\documentclass[]{iopart}
\usepackage{graphicx,amssymb,iopams}

\begin{document}
\title[Disks in a narrow channel jammed by gravity and centrifuge]{Disks in a
  narrow channel jammed by gravity and centrifuge: profiles of pressure, mass
  density and entropy density} \author{Christopher Moore$^{1}$, Dan
  Liu$^{2}$, Benjamin Ballnus$^{3}$, Michael Karbach$^{3}$, and Gerhard
  M{\"{u}}ller$^{1}$}

\address{$^{1}$ Department of Physics, University of Rhode Island, Kingston RI
  02881, USA} 

\address{$^{2}$ Department of Physics, University of New Haven, 
  West Haven CT 06516, USA} 

\address{$^{3}$ Bergische Universit{\"{a}}t Wuppertal,
  Fachbereich C, 42097 Wuppertal, Germany}


\begin{abstract}
  This work investigates jammed granular matter under conditions that produce
  heterogeneous mass distributions on a mesoscopic scale. We consider a system
  of identical disks that are confined 
  to a narrow channel, open at one end and closed off at the other end.  The
  disks are jammed by the local pressure in a gravitational field or
  centrifuge.  All surfaces are hard and frictionless.  We calculate the
  profiles of pressure, mass density, and entropy density on a mesoscopic
  length scale under the assumption that the jammed states are produced by
  random agitations of uniform intensity along the channel.  These profiles
  exhibit trends and features governed by the balancing of position-dependent
  forces and potential energies.  The analysis employs a method of
  configurational statistics that uses interlinking two-disk tiles as the
  fundamental degrees of freedom.  Configurational statistics weighs the
  probabilities of tiles according to competing potential energies associated
  with gravity and centrifugation.  Amendments account for the effects of the
  marginal stability of some tiles due to competing forces.
\end{abstract}


%
\section{Introduction}\label{sec:intro}
%
One criterion commonly used to demarcate the granular regime from the
colloidal regime in a suspension compares the energy $k_BT$ of thermal
fluctuations with the gravitational potential energy $mg\sigma$ of a suspended
particle when lifted a distance equal to its diameter $\sigma$ \cite{Fren11}.  It
follows that the earth's gravitational field $g$ has a part in terrestrial
experiments on granular matter. It can reasonably be argued that the
variations with vertical distance of quantities under investigations are
undetectably small in a wide range of experiments.  Effects of gravity are
real and interesting, nevertheless.  Jammed macrostates produced via some
protocol are described by characteristic profiles of pressure, mass density,
entropy density, mechanical coordination numbers etc.  Other fields or
potentials produce their own profiles in the same quantities.  We shall
consider combinations of gravity and centrifuge

How are trends and features in these profiles related to particular local
configurations of jammed grains?  Reliable answers to this wide-ranging
question are hard to get in general.  Evidence from experimental, theoretical,
and computational studies point to the balances of forces and potential
energies that determine the nature of jammed states [2-19].

We shall argue that the balance of relevant forces and the relative sizes of
relevant potential energies may affect the profiles in separate ways.  To
support this argument with evidence that is methodologically sound and, at the
same time, transparent regarding its physical origin, we consider a somewhat
idealized model: a system of disks confined to a narrow channel.

The work reported here builds on previous studies as described in the
following.  Ashwin, Bowles, and Saika-Voivod \cite{BS06, AB09, BA11}
introduced a simple model of rigid disks of one size and mass in a narrow
channel.  The disks are positioned in a horizontal plane and jammed by a
uniform pressure.  Two regimes of channel widths were identified, for which
the configurations of jammed disks can be described by a finite number of
interlinking tiles (four tiles in one regime and 32 tiles in the other).  The
configurational entropy versus volume fraction was calculated by elemenary
means in the first regime and by a transfer matrix technique in the second.

Gundlach \emph{et al.} \cite{GKLM13} generalized the analysis of the first
regime to a channel with its plane oriented vertically and its axis still
horizontal.  The effects of gravity were shown to produce a critical point at
which the system is highly susceptible to ordering tendencies.  Criticality
was shown to be robust in the presence of some effects of friction.
Remarkable analogies between this system of confined disks subject to gravity
and a system of hard spheres subject to friction \cite{SWM08, BSWM08, BSWM10,
  JM10, WSJM10, CPNC11} were identified, including the coexistence of domains
with different types or degrees of ordering.

Most recently, a system of disks confined to a narrow channel was used to
explore the relationship between thermally equilibrated fluid states and
mechanically stabilized jammed states, demonstrating new insights that remain
relevant for more complex systems \cite{YAB12,AYB13}.  The rigorous analysis
of idealized systems such as this often uncover issues and features that
transcend their own limitations.

\begin{figure}[t]
  \begin{center}
 \includegraphics[width=85mm]{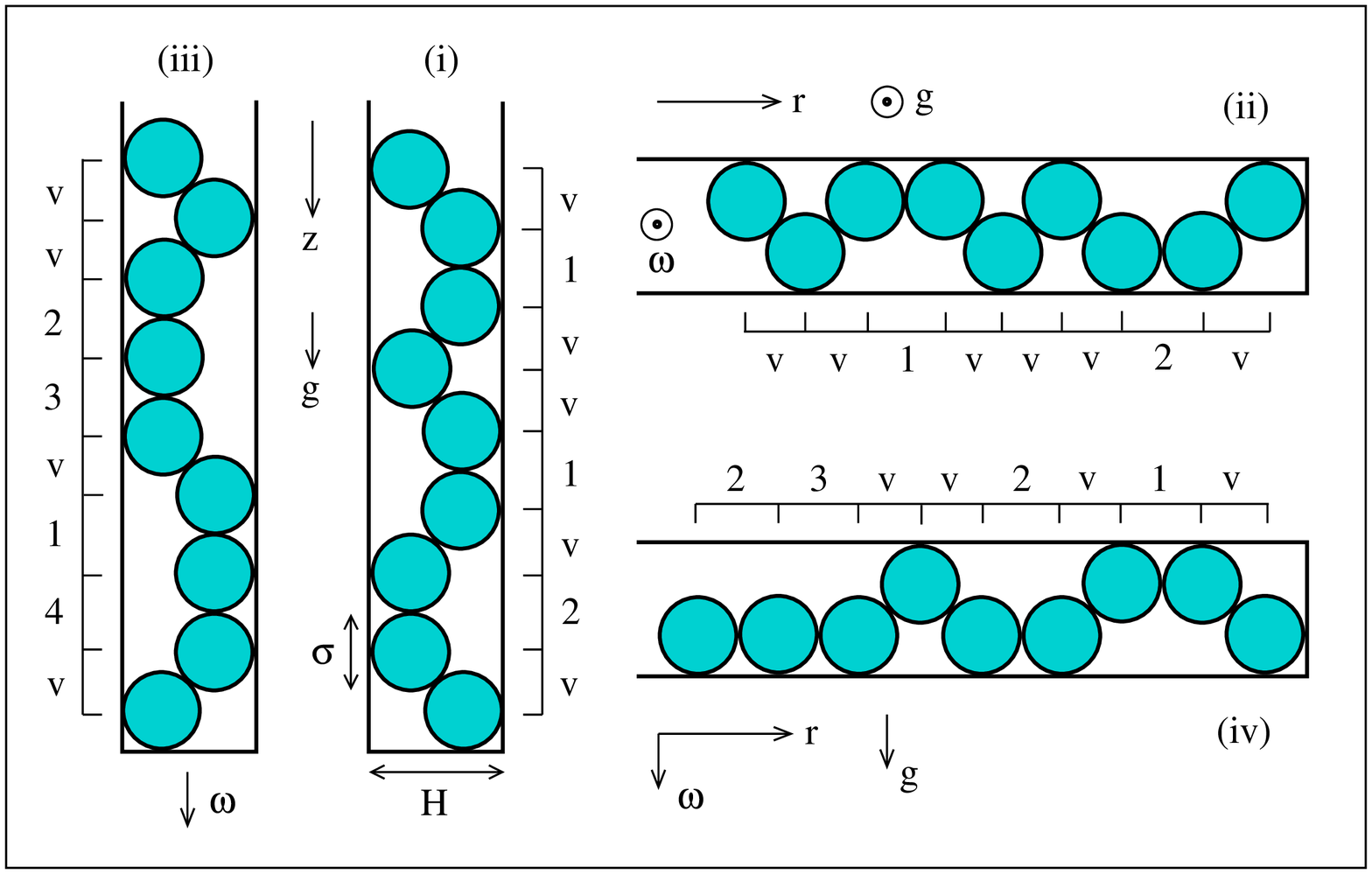}
\end{center}
\caption{Jammed microstates of disks of mass $\mu$ and diameter $\sigma$ in a
  channel of width $H$ for scenarios (i)-(iv).  For $H/\sigma\leq
  1+\sqrt{3/4}$ all configurations are sequences of interlinking two-disk
  tiles as shown.  In our methodology, tiles \textsf{v} are elements of
  pseudo-vacuum and tiles \textsf{1}, \textsf{2}, \textsf{3}, \textsf{4} are
  statistically interacting particles.}
  \label{fig:tiles}
\end{figure}

In this work we extend the analysis of disks jammed in a narrow channel to
conditions that produce heterogeneous mass distributions on a mesoscopic scale
caused by gravity and centrifuge (see Fig.~\ref{fig:tiles}).  We tilt the axis
of the channel into the (vertical) direction of the gravitational field
[scenario (i)] and spin the channel about its axis [scenario (iii)].  We
rotate the channel in a horizontal plane with its axis oriented radially and
its plane oriented horizontally [scenario (ii)] or vertically [scenario (iv)].
Previous studies of the statistical mechanics of jammed granular columns under
gravity in heterogeneous macrostates do exist [26-30] 
and will be discussed in Sec.~7.

The protocol of random agitations employed here is different from the ones
implied in Refs.~\cite{BS06, AB09, BA11, GKLM13}.  No piston is needed.
Jamming is caused by the gravitational force [scenario (i)], the centrifugal
force [scenario (ii)], or a combination of both [scenarios (iii) and (iv)]
acting on the disks when the random agitations are stopped suddenly.  These
forces vary along the channel, depending on the mass density, which itself
varies along the channel.  In scenarios (iii) and (iv) the mechanically stable
tiles sequences depend on the local jamming pressure.

Disks of mass $\mu$ and diameter $\sigma$ are confined to a long channel of
width $H$ with $H/\sigma= 1+\sqrt{3/4}$.  This particular width represents the
border between the aforementioned regimes \cite{AB09}.  All surfaces are rigid
and frictionless.  Jammed microstates are described as sequences of two-disk
tiles.  Successive tiles are interlinked by sharing one disk as illustrated in
Fig.~\ref{fig:tiles}.

Tiles come in four distinct shapes.  Two of them bear the same name
(\textsf{v}), the other two each bear two names, (\textsf{1,4}) and
(\textsf{2,3}), respectively, all for a reason.  The most compact state is an
alternating sequence of tiles \textsf{v}.  There is no need to assign
different names to the two shapes because all relevant attributes are
identical, yet they are not interchangeable.

In scenarios (i) and (ii) the least compact state is a sequence
$\mathsf{v\cdots1v2v1v2}\cdots\mathsf{v}$.  Here only tiles \textsf{v} can
follow each other directly.  Any tile \textsf{1} (\textsf{2}) is separated
from the next tile \textsf{1} \textsf({2}) by at least two tiles \textsf{v}.
A tile \textsf{1} is separated from the nearest tile \textsf{2} by at least
one tile \textsf{v}.

In scenario (iii) multiple tiles \textsf{1} and multiple tiles \textsf{2}
following one another directly can be stabilized by the centrifugal force
perpendicular to the channel axis at locations where the gravitational
pressure is not too strong.  Any tile \textsf{1} (\textsf{2}) that follows
directly another tile \textsf{1} (\textsf{2}) is renamed tile \textsf{4}
(\textsf{3}) in our analysis.

In scenario (iv) tiles \textsf{3} but not tiles \textsf{4} can be stabilized
by the gravitational force perpendicular to the channel axis at locations
where the centrifugal pressure is not too strong.  Conversely, below a certain
threshold centrifugal pressure tiles \textsf{v} and \textsf{1} are
destabilized by gravity in this scenario.

We describe our methodology in Sec.~\ref{sec:sit}.  The next four sections are
devoted each to one of the scenarios (i)-(iv) in ascending order of
complexity.

%
\section{Statistically interacting tiles}\label{sec:sit}
%
Every jammed microstate is a unique sequence of tiles subject to the rules
mentioned.  In the absence of acceleration due to gravity/rotation along the
channel, jammed macrostates are spatially homogeneous and are fully
characterized by the average populations of tiles from each species.

A detailed account of our methodology for such cases can be found in
Ref.~\cite{GKLM13}.  The method itself was invented by Haldane \cite{Hald91a},
and developed by Wu \cite{Wu94}, Isakov, \cite{Isak94}, Anghel \cite{Anghel},
and others \cite{LVP+08, copic, picnnn, pichs} in significant aspects.  The
adaptation to jammed granular matter as reported in Ref.~\cite{GKLM13} is
embedded in the framework of configurational statistics \cite{EO89, ME89}.

Here we extend that account to spatially heterogeneous situations such as
caused by gravitational or centrifugal forces.  Some aspects of this extension
are straightforward, others are more subtle.  Many conclusions reached in
Ref.~\cite{GKLM13} remain valid locally for mesoscopic segments of jammed
disks.

On any mesoscopic segment of channel we can interpret tiles \textsf{1},
\textsf{2}, \textsf{3}, \textsf{4} as statistically interacting quasiparticles
excited from the pseudo-vacuum $\mathsf{v}\mathsf{v}\cdots\mathsf{v}$ by
random agitations of given intensity in the face of a specific local pressure.
The number of distinct segments with given particle content is determined by
the multiplicity expression,
\begin{eqnarray}
  \label{eq:1a} 
  &W(\{N_m\}) =n_{pv}\prod_{m=1}^M
  \left(\begin{array}{c}d_m+N_m-1 \\ N_m\end{array}\right), 
  \\ \label{eq:1b} 
  &d_m =A_m-\sum_{m'=1} ^Mg_{mm'}(N_{m'}-\delta_{mm'}).
\end{eqnarray}
The pseudo-vacuum has degeneracy $n_{pv}=2$.  The number of particle species
is $M=4$.  In the taxonomy of Ref.~\cite{copic} they belong to the categories
host $(m=1,2)$ and tag $(m=3,4)$.  The capacity constants are
$A_1=A_2=\frac{1}{2}(N-1)$ and $A_3=A_4=0$, where $N$ is the number of disks.
The statistical interaction coefficients are compiled in Table~\ref{tab:1}.

\begin{table}[!h]\center{}
  \caption{Statistical interaction coefficients of tiles \textsf{1}, \textsf{2}, \textsf{3}, \textsf{4} in the role of quasiparticles.}\label{tab:1} 
    \begin{tabular}{c|rrrrr} \hline\hline  \rule[-2mm]{0mm}{6mm}
      $g_{mm'}$ & ~$1$ & ~~$2$ & ~~$3$ & ~~$4$ \\ \hline \rule[-2mm]{0mm}{6mm}
      $1$ & $\frac{3}{2}$ & $\frac{1}{2}$ & $\frac{1}{2}$  & $\frac{1}{2}$\\ \rule[-2mm]{0mm}{6mm}
      $2$ & $\frac{1}{2}$ & $\frac{3}{2}$ & $\frac{1}{2}$  & $\frac{1}{2}$ \\ \rule[-2mm]{0mm}{6mm}
      $3$ & $0$ & $-1$ & $0$ & $0$\\ \rule[-2mm]{0mm}{6mm}
      $4$ & $-1$ & $0$ & $0$ & $0$\\ \hline\hline 
    \end{tabular}
\end{table} 

The number of tiles is fixed at $N-1$.  Each \textsf{v}-tile contributes a
volume $\frac{1}{2}\sigma$ and each particle tile twice that.  Hence the
activation of a particle from any species expands the volume by
$\frac{1}{2}\sigma$.

Configurational statistics postulates that jammed macrostates with specific
profiles of mass density and entropy density can be generated reproducibly via
sufficiently well-defined protocols of random agitations.  Our calculations
assume a protocol in which the random agitations of controlled intensity are
applied uniformly along the channel.  When these agitations are stopped
abruptly, a jammed microstate is frozen out in short order by the local
gravitational or centrifugal pressure.

The associated jammed macrostate is then characterized by the (dimensionless) parameter field,
\begin{equation}\label{eq:2}
\hat{p}\doteq\frac{p(x)\sigma}{2T_k},
\end{equation}
where $p(x)$ is the local pressure and $T_k$ some measure for the intensity of
(uniform) random agitations.  We specify the position along the channel by
coordinate $x$ generically.  In applications we use $z$ if the axis is
vertical and $r$ if it is horizontal and rotating (see Fig.~\ref{fig:tiles}).

In scenarios (iii) and (iv) the jammed macrostate also depends on a second
(dimensionless) parameter, which is not a field:
\begin{equation}\label{eq:3} 
\hat{\omega}^2\doteq\frac{\mu\omega^2 d\sigma}{4T_k},\quad \hat{g}\doteq\frac{\mu gd}{T_k}, 
\end{equation}
respectively, where $d=\frac{1}{2}(H-\sigma)=(\sqrt{3}/4)\sigma$ is half the
shortest distance a disk can move between walls.  Moreover, in scenarios (i),
(iii) with gravitational pressure and (ii), (iv) with centrifugal pressure we
use, respectively, scaled position coordinates,
\begin{equation}\label{eq:47} 
\hat{z}\doteq\frac{2\mu gz}{T_k},\quad \hat{r}\doteq r\sqrt{\frac{\mu\omega^2}{T_k}}.
\end{equation}
Note that the $\omega$ in (\ref{eq:3}) describes a channel spinning about its
axis and the $\omega$ in (\ref{eq:47}) a channel with its axis rotating.

Key relations from the statistical mechanical analysis \cite{GKLM13} adapted
to heterogeneous situations are a set of linear equations for the scaled tile
population densities $\bar{N}_m(x)\doteq \langle N_m(x)\rangle/N$,
\begin{equation}\label{eq:4} 
w_m(x)\bar{N}_m(x)+\sum_{m'=1}^Mg_{mm'}\bar{N}_{m'}(x) =\bar{A}_m,
\end{equation}
where $\bar{A}_m\doteq A_m/N$, and where the $w_m(x)$ are real, non-negative
solutions of the nonlinear equations,
\begin{equation}\label{eq:5} 
e^{K_m(x)}=[1+w_m(x)]\prod_{m'=1}^M \big[1+w_{m'}(x)^{-1}\big]^{-g_{m'm}}.
\end{equation}
The $K_m(x)$ will be derived from (\ref{eq:2})-(\ref{eq:47}) case by case.
Our focus is on the profiles of the mass density,
\begin{equation}\label{eq:6} 
  \tilde{\rho}=\frac{\sigma}{2\mu}\rho(x)
  =\left[1+\sum_{m=1}^M\bar{N}_m(x)\right]^{-1},
\end{equation} 
and the entropy density,
\begin{eqnarray}\label{eq:7a}
\bar{S}(x) =& \sum_{m=1}^M 
\Big[\Big(\bar{N}_{m}(x)+\bar{Y}_m(x)\Big)\ln\Big(\bar{N}_m(x)+\bar{Y}_m(x)\Big)  
\nonumber \\
&  \hspace{15mm}-\bar{N}_m(x)\ln \bar{N}_m(x) -\bar{Y}_m(x)\ln \bar{Y}_m(x)\Big], 
\\ \label{eq:7b}
& \bar{Y}_m(x) \doteq \bar{A}_m-\sum_{m'=1}^Mg_{mm'} \bar{N}_{m'}(x),
\end{eqnarray}
in addition to that of pressure $p(x)$. One additional relation is necessary
for closure.  The local pressure (\ref{eq:2}), which enters (\ref{eq:5}) on
the left, depends on the profile of mass density (\ref{eq:6}) in a way that
varies between scenarios.  This relation will be introduced case by case.

%
\section{Gravity}\label{sec:grav}
%
In scenario (i) the axis of the channel is oriented vertically (see
Fig.~\ref{fig:tiles}).  Tiles \textsf{3} and \textsf{4} are mechanically
unstable against the slightest intensity of random agitations.  They do not
exist in jammed microstates produced by our protocol.  Configurational
statistics can accommodate this kind of mechanical instability by simply
freezing out tiles \textsf{3} and \textsf{4}.  Taking the limit $K_3,
K_4\to\infty$ leaves Eqs.~(\ref{eq:4}), (\ref{eq:5}), and
(\ref{eq:7a})-(\ref{eq:7b}) valid for tiles \textsf{1}{ and \textsf{2}.
A further simplification follows from symmetry: tiles \textsf{1} and
  \textsf{2} occur with equal probability.  We have
\begin{equation}\label{eq:8} 
K_1=K_2=\hat{p},
\end{equation}
in generalization of the zero-gravity situation \cite{GKLM13}.  Equations
(\ref{eq:5}) with (\ref{eq:8}) used on the left are valid on a mesoscopic
scale with the $z$-axis pointing vertically down and $z=0$ located at the top
of the pile of disks.  Anghel's rules \cite{Anghel} permit a type-1 merger
\cite{pichs} \textsf{1} \& \textsf{2} $\to$ $\mathsf{\bar{1}}$.  Symmetry
dictates that $w_1=w_2\doteq w$ and $\bar{N}_1=\bar{N}_2
\doteq\frac{1}{2}\bar{N}$.

The population profile of tiles \textsf{1}{ and \textsf{2} combined inferred
  from Eqs.~(\ref{eq:4}) and (\ref{eq:5}) depends on the pressure profile as
  follows:
\begin{equation}\label{eq:9}
\bar{N}=\big[w+2\big]^{-1}, \quad w=\frac{1}{2}e^{\hat{p}}\Big[1+\sqrt{1+4e^{-\hat{p}}}\Big].
\end{equation}
The local pressure is caused by the weight of the disks piled on top.  It
depends on the mass density (\ref{eq:6}) transcribed into
\begin{equation}\label{eq:10}
  \tilde{\rho}=\frac{1}{1+\bar{N}},
\end{equation}
via the integral,
\begin{equation}\label{eq:11}
p(z)=g\int_0^z dz'\rho(z'),
\end{equation}
which we invert into
\begin{equation}\label{eq:48} 
\hat{z}=2\int_0^{\hat{p}} d\hat{p}'\left[1+\bar{N}(\hat{p}')\right].
\end{equation}
The integral yields
\begin{equation}\label{eq:12}
  e^{\hat{z}} 
  = e^{3\hat{p}} \frac{\sqrt{5}+1}{\sqrt{5}-1}
  \frac{\sqrt{1+4e^{-\hat{p}}}-1}{\sqrt{1+4e^{-\hat{p}}}+1}.
\end{equation}
The entropy density (\ref{eq:7a})-(\ref{eq:7b}) for this case becomes
\begin{eqnarray}\label{eq:13}
\bar{S}=\Big[\big(1-\bar{N}\big)\ln\big(1-\bar{N}\big)
-\bar{N}\ln\bar{N} 
-\big(1-2\bar{N}\big)\ln\big(1-2\bar{N}\big)\Big].
\end{eqnarray}
Expressing (\ref{eq:12}) as a cubic equation,
\begin{equation}\label{eq:15} 
(h^2-1)(h+1)=4(1+\sqrt{5})e^{-\hat{z}/2},
\end{equation}
in $h(\hat{p})\doteq\sqrt{1+4e^{-\hat{p}}}$, we find a unique real solution
with asymptotics,
\begin{equation}\label{eq:19} 
  h = \left\{
  \begin{array}{ll}\displaystyle\label{eq:30}
    \sqrt{5} - \frac{1+3\sqrt{5}}{22}\hat{z} + \mathrm{O}(\hat{z}^{2}),
    & \hat{z} \ll 1 \\ \displaystyle
    1 +\big(1+\sqrt{5}\big)e^{-\hat{z}/2}+ \mathrm{O}(e^{-\hat{z}}), 
    & \hat{z} \gg 1 \rule[-2mm]{0mm}{6mm}
  \end{array}\right. .
\end{equation}

The curve of $\hat{p}-\frac{1}{2}\hat{z}$ versus $\hat{z}$ in
Fig.~\ref{fig:scen1} represents the (negative) deviation from hydrostatic
pressure of close-packed disks.  Also shown are the pressure asymptotes near
the top of the stack and deep down in the channel.

\begin{figure}[ht]
  \begin{center}
 \includegraphics[width=52mm]{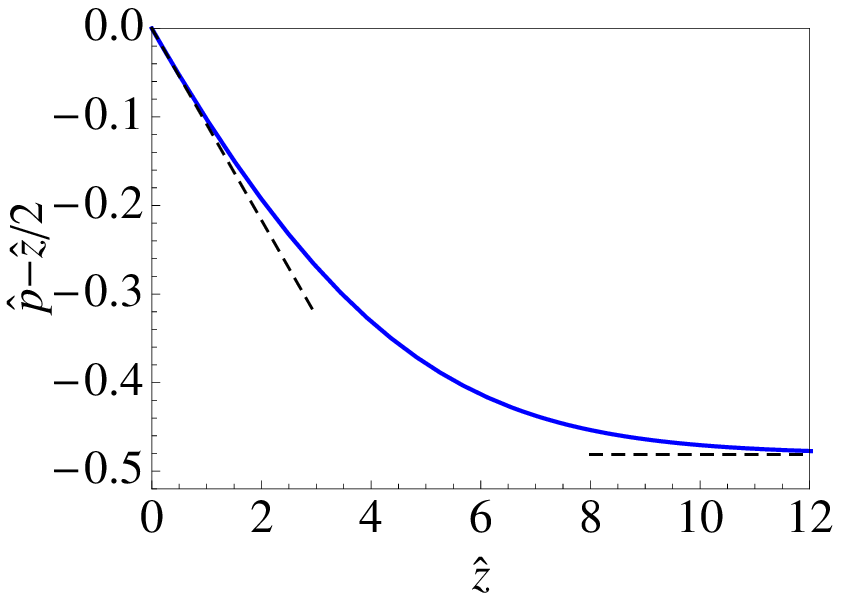}
 \includegraphics[width=52mm]{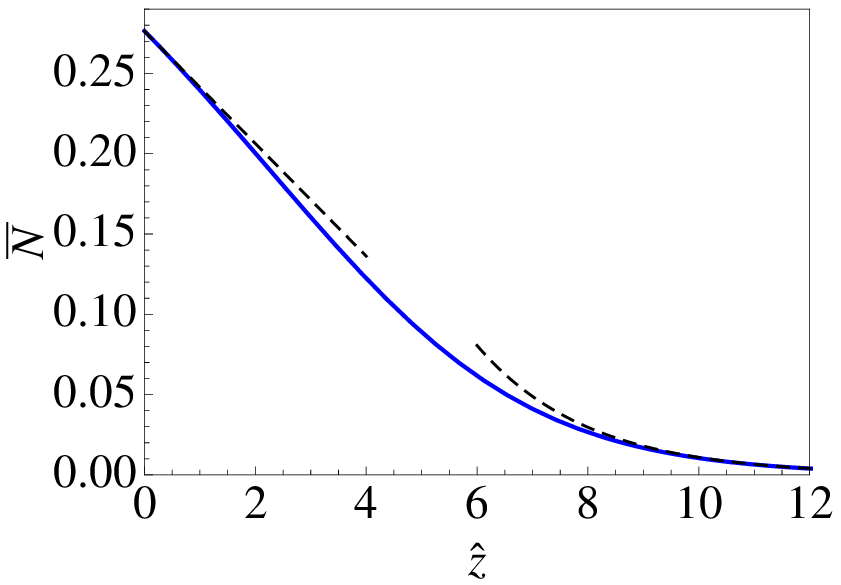}

  \includegraphics[width=52mm]{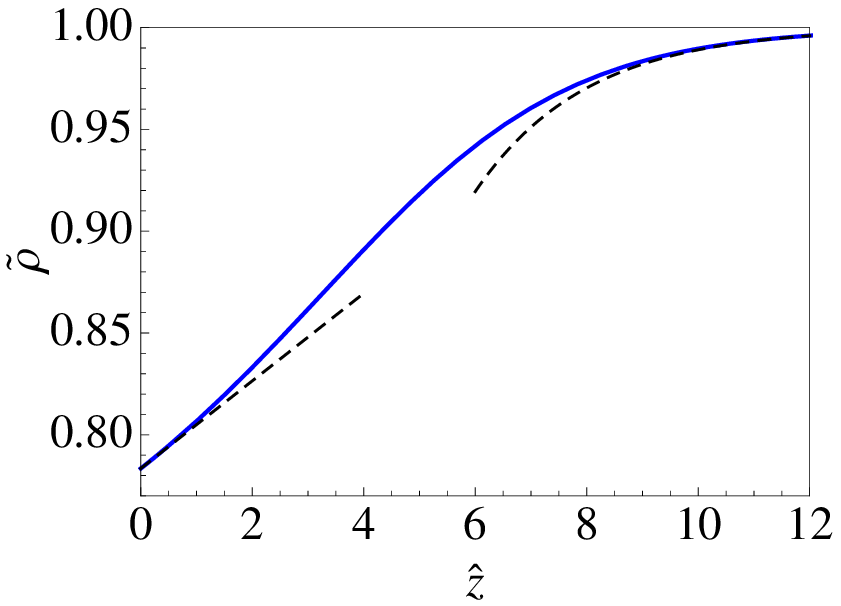}\hspace{2mm}%
  \includegraphics[width=52mm]{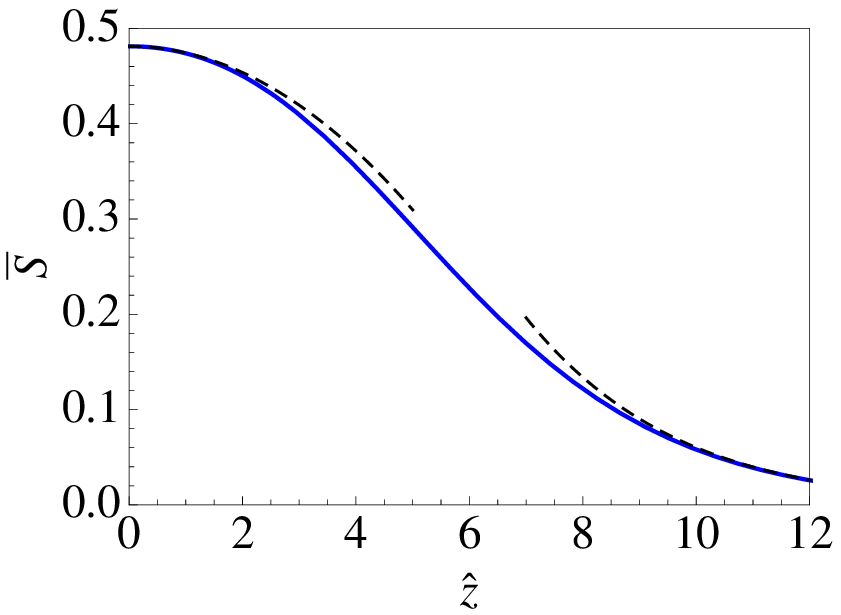}
\end{center}
\caption{Profiles of scaled pressure, $\hat{p}$, scaled population density of
  particle tiles $\bar{N}$, scaled mass density $\tilde{\rho}$, and scaled entropy
  $\bar{S}$ versus scaled depth $\hat{z}$ in scenario (i). The asymptotics are
  shown dashed.}
  \label{fig:scen1}
\end{figure}

The pressure has a universal profile in the scaled variables with asymptotics,
\begin{equation}\label{eq:16} 
\hat{p}=\left\{\begin{array}{ll}
 {\displaystyle   \frac{15+\sqrt{5}}{44}\hat{z} + \mathrm{O}(\hat{z}^{2})}, & \hat{z}\ll1 \\
 {\displaystyle  \frac{\hat{z}}{2} -\mathrm{Arcoth}\sqrt{5} + \mathrm{O}(e^{-\hat{z}/2})}, & \hat{z}\gg1
  \end{array} \right..
\end{equation}
The actual pressure $p$ at actual depth $z$ does, of course, depend on the
intensity $T_k$ of random agitations used in our jamming protocol.  The
overall pressure deep down decreases with increasing $T_k$ as expected.

That same solution enters the expression for the particle population density
and the mass density as follows:
\begin{equation}\label{eq:17} 
 \bar{N} = \frac{h(\hat{p})-1}{2h(\hat{p})},\quad   
 \tilde{\rho} = \frac{2h(\hat{p})}{3h(\hat{p})-1}.
\end{equation}
The variation of both densities versus scaled depth as depicted in
Fig.~\ref{fig:scen1} is linear near the top and exponential deep down.  The
asymptotics are
\begin{equation}
  \label{eq:18}
  \tilde{\rho}= \left\{
  \begin{array}{ll} \displaystyle
    \frac{2\sqrt{5}}{3\sqrt{5}-1} 
    + \frac{4}{(3\sqrt{5}-1)^{3}}\hat{z} + \mathrm{O}(\hat{z}^{2}),
    & \hat{z} \ll 1 \\ \displaystyle
    1-\frac{\sqrt{5}+1}{2}e^{-\hat{z}/2}+ \mathrm{O}(e^{-\hat{z}}), 
    & \hat{z} \gg 1
  \end{array}\right. ,
\end{equation}
\begin{equation}
  \label{eq:33}
  \bar{N} = \left\{
  \begin{array}{ll} \displaystyle
    \frac{\sqrt{5}-1}{2} - \frac{1+3\sqrt{5}}{5\cdot 44}\hat{z} + \mathrm{O}(\hat{z}^{2}),
    & \hat{z} \ll 1 \\ \displaystyle
    \frac{\sqrt{5}+1}{2}e^{-\hat{z}/2}+ \mathrm{O}(e^{-\hat{z}}), 
    & \hat{z} \gg 1 \rule[-2mm]{0mm}{8mm}
  \end{array} \right. ,
\end{equation}
The rate at which $\tilde{\rho}$ grows with depth first increases then
decreases toward zero.

The profile of the entropy density inferred from (\ref{eq:13}), (\ref{eq:17}),
and the physical solution of (\ref{eq:15}) has the compact form
\begin{eqnarray}\label{eq:21}
  \bar{S} = \mathrm{Arsinh}\left(\frac{1}{2}e^{\hat{p}/2}\right) - \frac{\hat{p}}{2h(\hat{p})}.
\end{eqnarray}
The asymptotic expansions in terms of $\tilde{z}$ are
\begin{equation}\label{eq:22}
  \bar{S}= \left\{
  \begin{array}{ll} \displaystyle
    \ln\frac{1+\sqrt{5}}{2} -
    \left(\frac{1+3\sqrt{5}}{44}\right)^{2}\frac{\hat{z}^{2} }{2\sqrt{5}}
    + O(\tilde{z}^{3}),
    & \hat{z} \ll 1 
    \\ \displaystyle
    \frac{1+\sqrt{5}}{2}
    \left(\frac{\hat{z}}{2}+1-\ln\frac{1+\sqrt{5}}{2}\right)e^{-\hat{z}/2}
    + O(e^{-\hat{z}}),
    & \hat{z} \gg 1.
  \end{array}\right.
\end{equation}
It is a monotonically
decreasing function as shown in Fig.~\ref{fig:scen1}, with a quadratic
dependence on $\hat{z}$ near the top and an exponential tail deep down.

Spinning the vertical channel about its axis stabilizes tiles \textsf{3} and
\textsf{4} at sufficiently low gravitational pressure.  How this affects the
four profiles of Fig.~\ref{fig:scen1} will be discussed in
Sec.~\ref{sec:grav+cent}.

%
\section{Centrifuge}\label{sec:cent}
%
Scenario (ii) considers a channel with plane and axis oriented horizontally,
rotating with constant angular velocity $\omega$ in the horizontal plane (see
Fig.~\ref{fig:tiles}).  All densities are now functions of the radial
coordinate $r$.  The disks are jammed by the local pressure due to
centrifugation.

Only tiles \textsf{1} and \textsf{2} are stable and the same symmetries obtain
as in scenario (i).  Equations~(\ref{eq:8})-(\ref{eq:10}) and (\ref{eq:13})
remain valid whereas Eqs.~(\ref{eq:11}) and (\ref{eq:48}) are to be replaced,
respectively, by
\begin{eqnarray}\label{eq:20} 
& p(r)=\omega^2\int_0^r dr'r'\rho(r'),\\ \label{eq:14}
& \hat{r}^2=2\int_0^{\hat{p}} d\hat{p}'\left[1+\bar{N}(\hat{p}')\right].
\end{eqnarray}
We assume that the column of disks just reaches the location of the axis of rotation.
This is a convenient albeit somewhat artificial boundary condition.
It preserves all features of interest and avoids clutter.
The resulting centrifugal pressure profile reads
\begin{equation}\label{eq:35}
  e^{\hat{r}^2} 
  = e^{3\hat{p}} \frac{\sqrt{5}+1}{\sqrt{5}-1}
  \frac{\sqrt{1+4e^{-\hat{p}}}-1}{\sqrt{1+4e^{-\hat{p}}}+1}.
\end{equation}

The four profiles shown in Fig.~\ref{fig:scen2} differ from those shown in
Fig.~\ref{fig:scen1} in ways that are as significant physically as they are
obvious mathematically.  Linear (quadratic) dependences at small $\hat{z}$ are
replaced by quadratic (quartic) dependences at small $\hat{r}$.  Exponential
behavior at large $\hat{z}$ is replaced by Gaussian behavior at large
$\hat{r}$.

\begin{figure}[htb]
  \begin{center}
 \includegraphics[width=52mm]{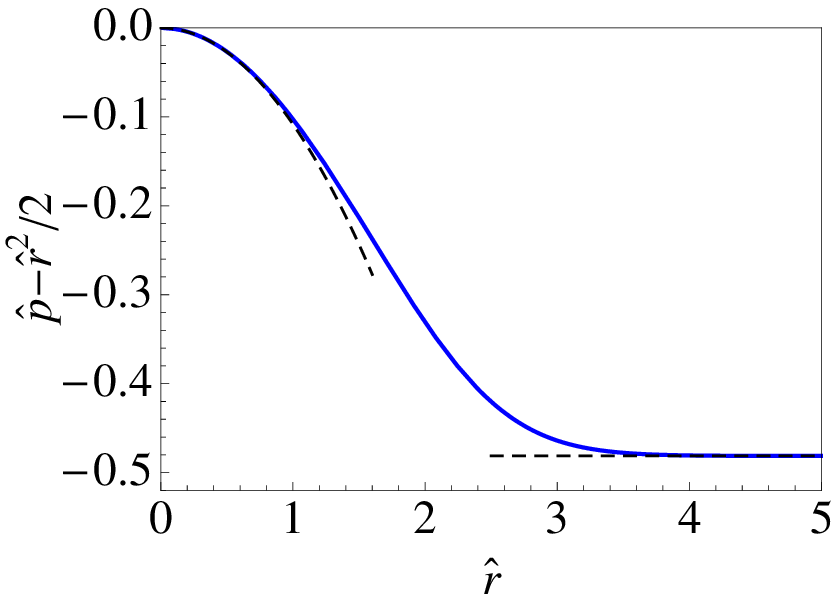}\hspace{2mm}%
 \includegraphics[width=52mm]{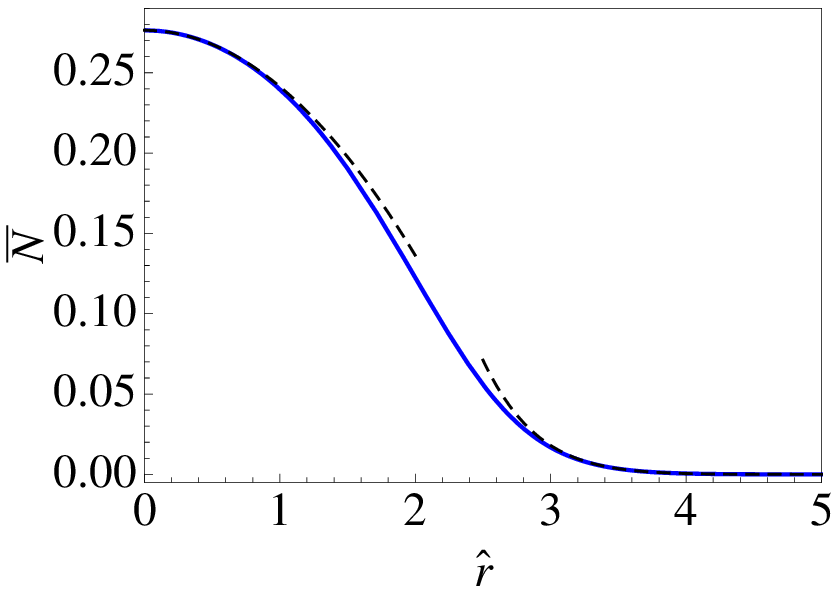}

  \includegraphics[width=52mm]{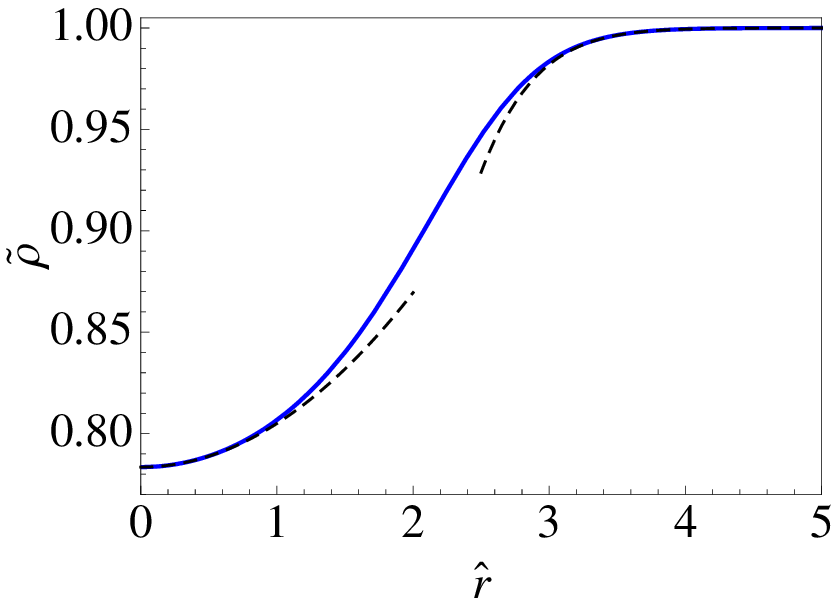}\hspace{2mm}%
  \includegraphics[width=52mm]{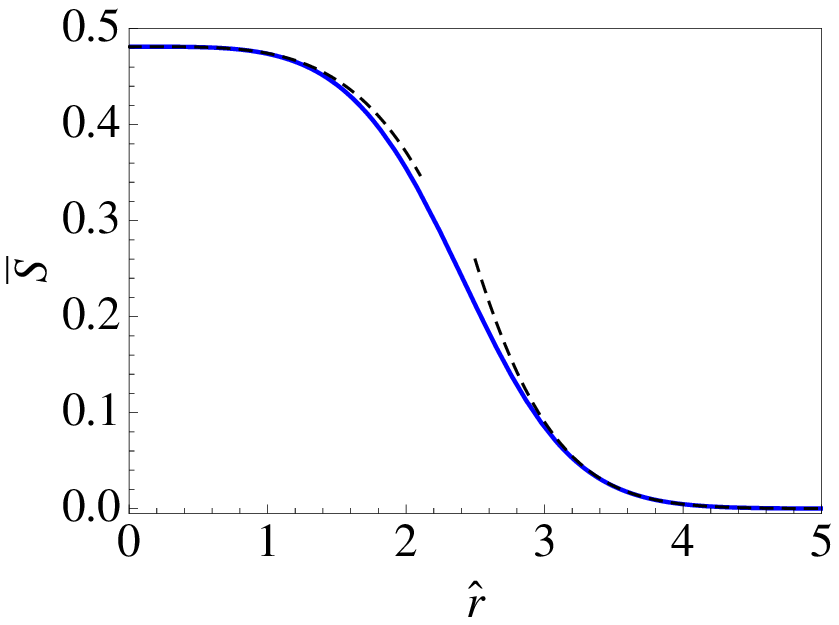}
\end{center}
\caption{Profiles of scaled pressure $\hat{p}$, scaled population density of
  particle tiles $\bar{N}$, scaled mass density $\tilde{\rho}$, and scaled
  entropy $\bar{S}$ versus scaled radius $\hat{r}$ in scenario (ii). The
  asymptotics are shown dashed.}
  \label{fig:scen2}
\end{figure}

These simple and smooth profiles for a rotating channel will acquire
additional structures when the plane is turned from a horizontal into a
vertical direction.  In that case the mechanical stability of some tiles
depends on the centrifugal pressure.  The four profiles thus modified will be
discussed in Sec.~\ref{sec:cent+grav}.

%
\section{Gravity with centrifuge}\label{sec:grav+cent}
%
Scenario (iii) is an extension of scenario (i).  The channel is spinning about
its vertical axis with constant angular velocity $\omega$ (see
Fig.~\ref{fig:tiles}).  The spinning has the effect of mechanically
stabilizing tiles \textsf{3} and \textsf{4} provided the local gravitational
pressure is not too high.

In all jammed states every disk touches a wall.  Hence each disk experiences
the same centrifugal force irrespective of what tile it is part of.  Only
disks that have both neighbors touching the wall on their side need the
centrifugal force for their own mechanical stability.  All such disks belong
to a tile \textsf{3} or \textsf{4}.

Tiles \textsf{v}, \textsf{1}, \textsf{2} are stable for any pressure.  The
stability of tiles \textsf{3} and \textsf{4} is precarious at all nonzero
pressures but to varying degrees.  Near the top there exists an asymptotic
low-pressure regime where tiles \textsf{3} and \textsf{4} are no less stable
than tiles \textsf{1} and \textsf{2}.  Deep down there exists an asymptotic
high-pressure regime where tiles \textsf{3} and \textsf{4} are highly
unstable.  The physics of the two asymptotic regimes is firmly grounded in
configurational statistics.

\subsection{Low pressure and high pressure}\label{sec:top}

The analysis for the low-pressure asymptotic regime proceeds parallel to that
of Sec.~\ref{sec:grav} with some modifications.  We must replace
Eq.~(\ref{eq:8}) by
\begin{equation}\label{eq:23} 
K_m=\hat{p},\quad m=1,\ldots,4.
\end{equation}
Before we solve Eqs.~(\ref{eq:4}) and (\ref{eq:5}) for $M=4$ statistically
interacting particles we take advantage of the symmetries reflected in
(\ref{eq:23}) and in the $A_m, g_{mm'}$ from Table~\ref{tab:1}.  According to
Anghel's rules \cite{Anghel, pichs} we can implement three successive mergers,
\begin{equation}\label{eq:37} 
 \mathsf{3} ~\&~ \mathsf{4} \to \mathsf{\bar{3}},
 \quad \mathsf{1} ~\&~ \mathsf{2} \to \mathsf{\bar{1}},\quad
 \mathsf{\bar{1}} ~\&~ \mathsf{\bar{3}} \to \mathsf{\tilde{1}},
\end{equation}
that lead to a single species of fermionic particles: 
\begin{equation}\label{eq:24} 
K_{\tilde{1}}=\hat{p},\quad A_{\tilde{1}}=N-1,\quad g_{\tilde{1}\tilde{1}}=1.
\end{equation}
The dependence on the pressure profile of the population profile for particles
$\tilde{1}$ inferred from (\ref{eq:4}) then becomes
\begin{equation}\label{eq:25} 
\bar{N}=\frac{1}{e^{\hat{p}}+1}.
\end{equation}
With (\ref{eq:25}) substituted into (\ref{eq:10}), the solution of (\ref{eq:11}) is
\begin{equation}\label{eq:26} 
\frac{\hat{z}}{2}=\hat{p}+\ln\left(\frac{2}{1+e^{-\hat{p}}}\right),
\end{equation}
in analogy to (\ref{eq:12}) but solved more compactly:
\begin{equation}\label{eq:27} 
\hat{p}=\frac{\hat{z}}{2}-\ln\left(\frac{4}{1+\sqrt{1+8e^{-\hat{z}/2}}}\right).
\end{equation}
The configurational entropy density derived from (\ref{eq:7a}-\ref{eq:7b}) is
recognizably fermionic:
\begin{eqnarray}\label{eq:28} 
\bar{S} =-\bar{N}\ln\bar{N}-\big(1-\bar{N}\big)\ln\big(1-\bar{N}\big) 
 =\frac{\hat{p}}{1+e^{\hat{p}}}+\ln\left(1+e^{-\hat{p}}\right).
\end{eqnarray}

The four profiles predicted by configurational statistics for the low-pressure
asymptotic regimes are shown in Fig.~\ref{fig:scen3as} as the set of curves
drawn solid at small $\hat{z}$ and dashed at large $\hat{z}$.  The underlying
assumption that tiles \textsf{3} and \textsf{4} are mechanically stable is
accurate near the surface but invalid deep down.

\begin{figure}[!h]
  \begin{center}
 \includegraphics[width=52mm]{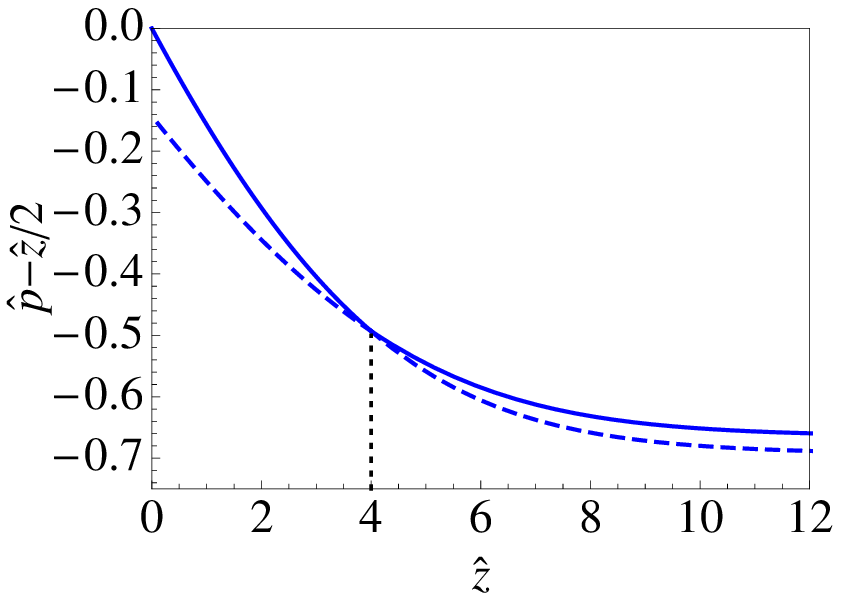}\hspace{2mm}%
 \includegraphics[width=52mm]{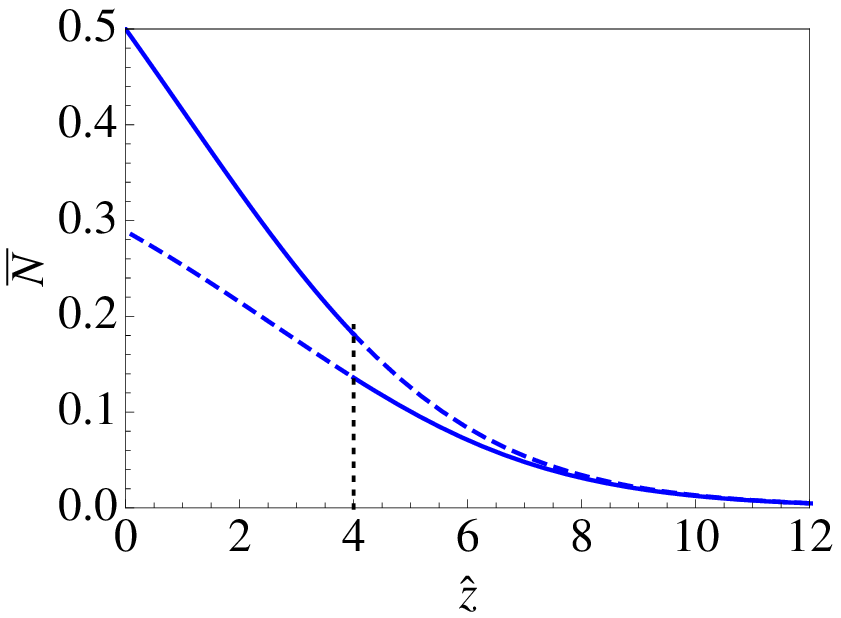}
  \includegraphics[width=52mm]{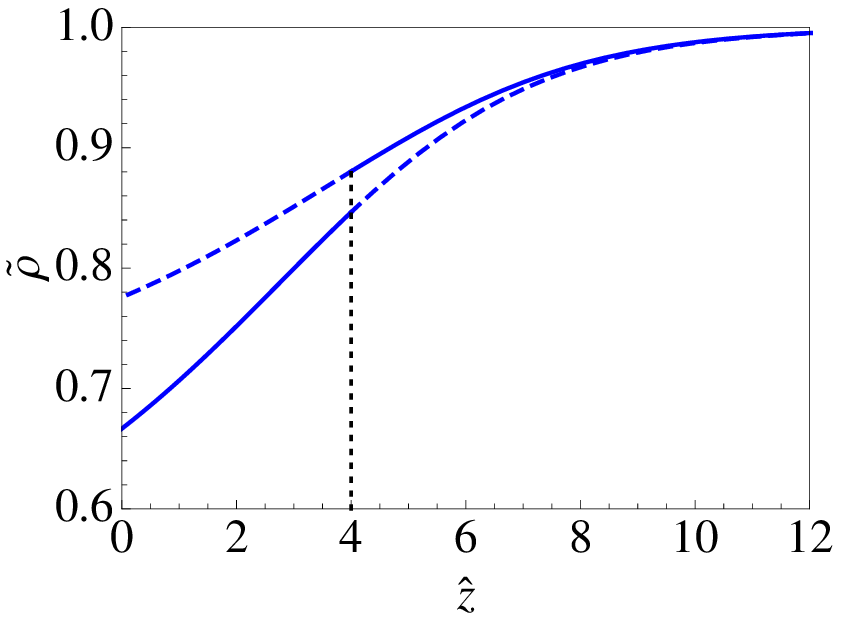}\hspace{2mm}%
  \includegraphics[width=52mm]{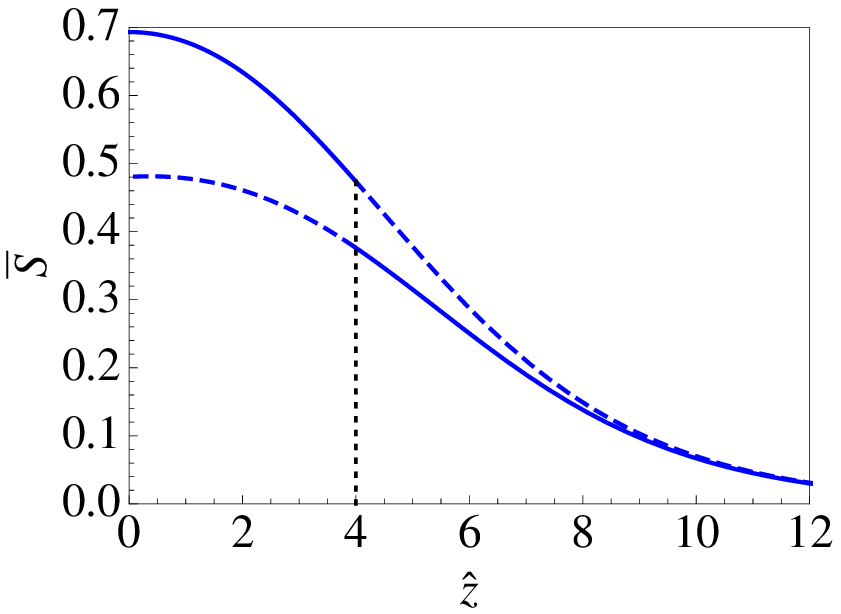}
\end{center}
\caption{Profiles of scaled pressure, scaled population density of particle
  tiles, scaled mass density, and scaled entropy, versus scaled depth in the
  low-pressure asymptotic regime (solid at $\hat{z}<4$, dashed at $\hat{z}>4$)
  and in the high-pressure asymptotic regime (dashed at $\hat{z}<4$, solid at
  $\hat{z}>4$. }
  \label{fig:scen3as}
\end{figure}

At $\hat{z}\gg1$ the high gravitational pressure destabilizes practically all
tiles \textsf{3} and \textsf{4} under jamming.  This high-pressure asymptotic
regime leads us back to the statistical mechanics of scenario (i).  If we
assume that all tiles \textsf{3} and \textsf{4} lose mechanical stability at
the same threshold pressure $p_0$, then we can use Eq. (\ref{eq:11}) modified
into
\begin{equation}\label{eq:39}
p(z)-p_0=g\int_{z_0}^z dz'\rho(z').
\end{equation}
The solution for scaled variables is
\begin{equation}\label{eq:29} 
e^{\hat{z}-\hat{z}_0}=e^{3(\hat{p}-\hat{p}_0)}
\frac{2+e^{\hat{p}_0}\big[1+\sqrt{1+4e^{-\hat{p}_0}}\big]}
{2+e^{\hat{p}}\big[1+\sqrt{1+4e^{-\hat{p}}}\big]}.
\end{equation}
We arrive at the four high-pressure asymptotic profiles shown in
Fig.~\ref{fig:scen3as} as the set of curves drawn dashed at low $\hat{z}$ and
solid at large $\hat{z}$.  The values $\hat{z}_0$ and $\hat{p}_0$ are related
by the requirement that the pressure be continuous across the point of
mechanical instability.  For the purpose of illustration we choose
$\hat{z}_0=4$, implying $\hat{p}_0\simeq1.51$,

This oversimplified model of mechanical instability predicts a kink in the
pressure profile and discontinuities in the other three profiles.  In reality,
the mechanical instability of tiles \textsf{3} and \textsf{4} is a more
complex phenomenon as will be discussed next.

\subsection{From low to high pressure}\label{sec:middle}
Tiles \textsf{3} and \textsf{4} have conditional mechanical stability.  Only a
fraction of such tiles that are predicted by configurational statistics to be
generated survive jamming.  That fraction becomes smaller with increasing
pressure.

To describe the physics of this situation most economically we perform the
first two mergers (\ref{eq:37}) of particle tiles.  This leads to two species
$\mathsf{\bar{1}}$ and $\mathsf{\bar{3}}$ with specifications
$A_{\bar{1}}=N-1$, $A_{\bar{3}}=0$, $g_{\bar{1}\bar{1}}=2$,
$g_{\bar{1}\bar{3}}=1$, $g_{\bar{3}\bar{1}}=-1$, and $g_{\bar{3}\bar{3}}=0$.
Tiles \textsf{1} and \textsf{2} are represented by particles
$\mathsf{\bar{1}}$ and tiles \textsf{3} and \textsf{4} by particles
$\mathsf{\bar{3}}$.

The population densities of particles $\mathsf{\bar{1}}$ and
$\mathsf{\bar{3}}$ created under local gravitational pressure $\hat{p}$,
\begin{equation}\label{eq:32} 
\bar{N}_{\bar{1}}(\hat{p})=\frac{e^{\hat{p}}}{\big(1+e^{\hat{p}}\big)^2},\quad 
\bar{N}_{\bar{3}}(\hat{p})=\frac{1}{\big(1+e^{\hat{p}}\big)^2},
\end{equation}
are independent of $\hat{\omega}$ because each disk experiences the same
centrifugal force.

The ratio $\bar{N}_{\bar{3}}/\bar{N}_{\bar{1}}=e^{-\hat{p}}$ tells us that
particles $\mathsf{\bar{3}}$ with precarious mechanical stability are already
created in lower proportions than the mechanically stable particles
$\mathsf{\bar{1}}$.  That ratio is close to unity at low pressure (near the
top) and becomes exponentially small at high pressure (deep down).  In fact,
both population densities become small at high pressure and the
\textsf{v}-tiles become predominant.

In \ref{sec:scen-iii} we present a model for the conditional
mechanical stability of particles $\mathsf{\bar{3}}$.  It predicts that of the
$N_{\mathsf{\bar{3}}}$ particles $\mathsf{\bar{3}}$ generated at angular
velocity $\hat{\omega}$ the fraction that survives jamming under gravitational
pressure $\hat{p}$ is
\begin{equation}\label{eq:36} 
R_\omega(\hat{p}/\hat{\omega}^2)=\frac{\sigma/d}{\sigma/d+\hat{p}/\hat{\omega}^2}.
\end{equation}
The limits $R_\omega(0)=1$ and $R_\omega(\infty)=0$ represent the physics in
the low-pressure and high-pressure asymptotic regimes, respectively.

The total population density of particles $\mathsf{\bar{1}}$ and
$\mathsf{\bar{3}}$ at scaled gravitational pressure $\hat{p}$ and scaled
angular velocity $\hat{\omega}$ that survive jamming thus becomes
\begin{eqnarray}\label{eq:31} 
\bar{N}^{(c)}(\hat{p},\hat{\omega}) \doteq\bar{N}_{\bar{1}}(\hat{p})
+R_\omega(\hat{p}/\hat{\omega}^2)\bar{N}_{\bar{3}}(\hat{p}) 
=\frac{R_\omega(\hat{p}/\hat{\omega}^2)+e^{\hat{p}}}{\big(1+e^{\hat{p}}\big)^2},
\end{eqnarray}
which approaches (\ref{eq:25}) at low $\hat{p}/\hat{\omega}^2$ and
(\ref{eq:9}) at high $\hat{p}/\hat{\omega}^2$ as required for consistency.
Relation (\ref{eq:48}) now holds for $\hat{z}(\hat{p},\hat{\omega})$ and
$\bar{N}^{(c)}(\hat{p},\hat{\omega})$.

\begin{figure}[b]
  \begin{center}
 \includegraphics[width=52mm]{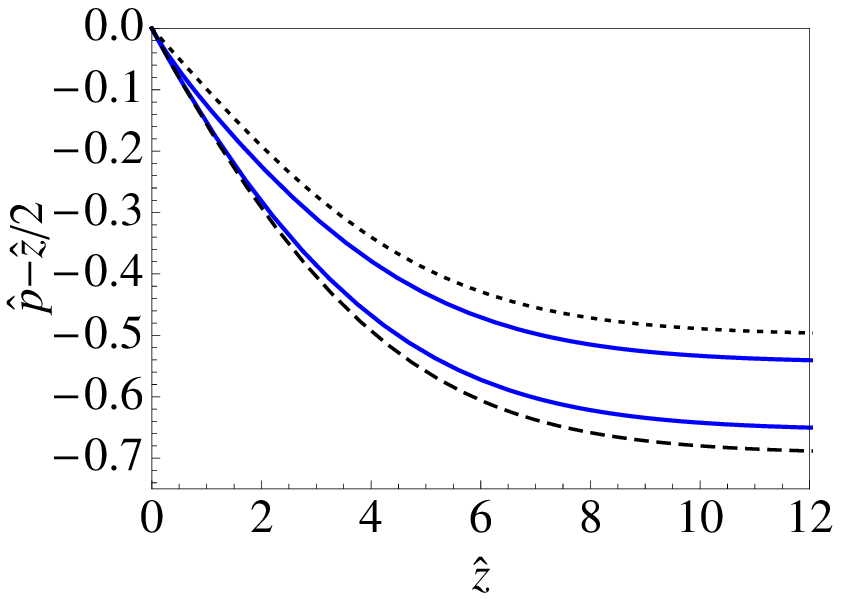}\hspace{2mm}%
 \includegraphics[width=52mm]{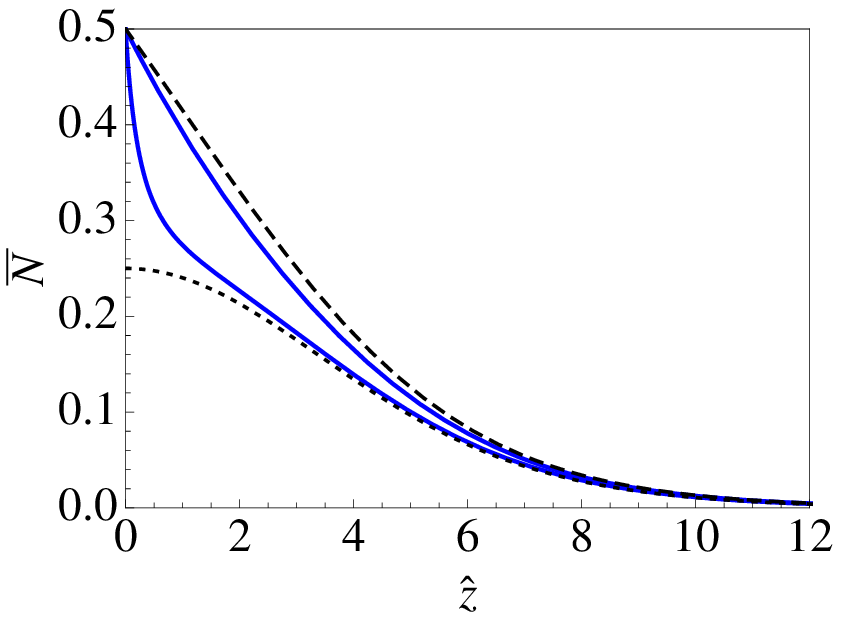}
  \includegraphics[width=52mm]{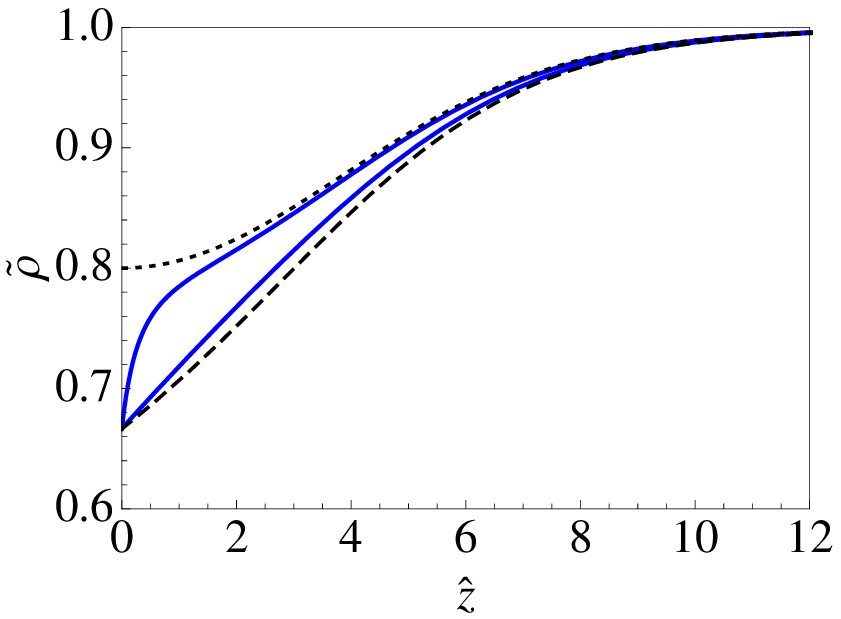}\hspace{2mm}%
  \includegraphics[width=52mm]{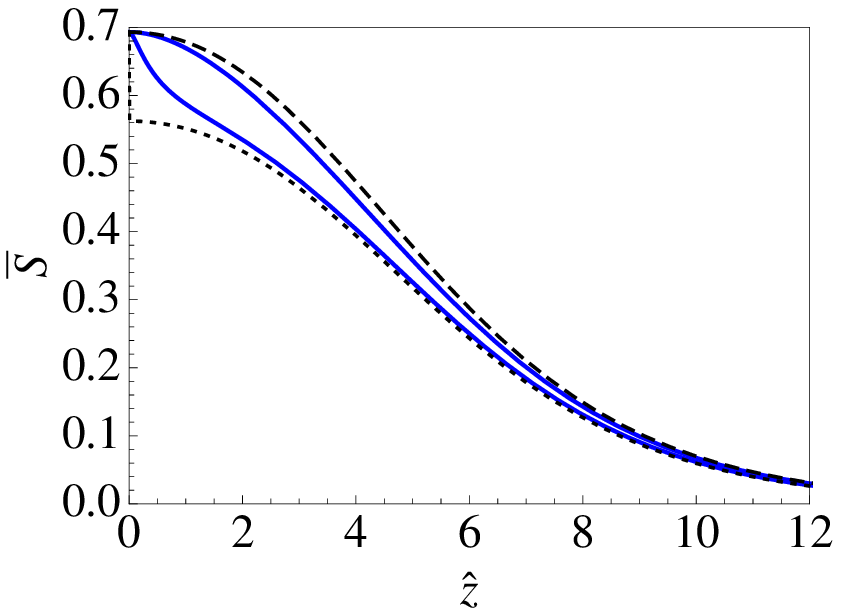}
\end{center}
\caption{Profiles of scaled pressure, scaled population density of particle
  tiles, scaled mass density, and scaled entropy, versus scaled depth for
  scaled angular velocities $\hat{\omega}=\infty$ (dashed line),
  $\hat{\omega}=1, 0.2$ (solid lines), and $\hat{\omega}=0$ (dotted line).}
  \label{fig:scen3lh}
\end{figure}

The four profiles emerging from this more realistic model for the conditional
stability of particles $\mathsf{\bar{3}}$, now depend on the parameter
$\hat{\omega}$, which compares the centrifugal energy of the disks in the
spinning channel with the intensity $T_k$ of random agitations used in the
jamming protocol.  These profiles are shown in the four panels of
Fig.~\ref{fig:scen3lh}.  All curves are smooth.  The singularities predicted
by the more primitive model (Fig.~\ref{fig:scen3as}) have disappeared.

The effects of centrifugal force on the disks in the spinning channel are manifest in the deviations from the doted curve shown in each panel.
Faster spinning stabilizes an increasing number of particles $\mathsf{\bar{3}}$. 
Hence $\bar{N}$ is enhanced.
Additional particles produce more disorder, which, in turn, increases $\bar{S}$.
The same particles added, on the other hand, suppress the mass density $\rho$ at the location $\hat{z}$ where they are placed and thus reduce the pressure $\hat{p}$ at any distance $\hat{z}$ measured from the top.

%
\section{Centrifuge with gravity}\label{sec:cent+grav}
%
Physically, scenario (iv) is an extension of scenario (ii).  The axis of the
channel is horizontal again.  Rotation produces centrifugal pressure.  The
plane of the channel is now vertical with gravity pulling the disks toward the
lower wall (see Fig.~\ref{fig:tiles}).

Mathematically, scenario (iv) is akin to scenario (iii) but more complex.
Only one of the five tiles is mechanically stable at all pressures.  The
strengths and limitations of configurational statistics are most clearly
brought to light in this scenario.

The four familiar profiles depend separately on $\hat{p}$ and $\hat{g}$, not
merely on their ratio.  Therefore, the results for light disks and heavy disks
cannot be reduced to the same universal curves via scaling.  The attributes
`light' $(\hat{g}\ll1)$ and `heavy' $(\hat{g}\gg1)$ are gauged by the ratio of
the work needed to lift one disk halfway across the channel in units of the
intensity of random agitations.

Tiles \textsf{2} are mechanically stable and tiles \textsf{4} unstable at any
pressure.  In \ref{sec:scen-iv} we show that tiles \textsf{3} are
mechanically stable up to pressure $\hat{p}=\frac{1}{3}\hat{g}$.  At higher
pressure they are only conditionally stable in the sense that only a fraction
of tiles \textsf{3} generated by random agitations survive.  Conversely, tiles
\textsf{v} are unstable at $\hat{p}\leq\frac{1}{3}\hat{g}$ and conditionally
stable at $\hat{p}>\frac{1}{3}\hat{g}$.  Tiles \textsf{1}, by contrast, remain
unstable up to the higher threshold pressure, $\hat{p}=\frac{2}{3}\hat{g}$,
and then become conditionally stable.

For a proper understanding of the physics in this scenario, it is important
that we distinguish the balancing of potential energies and the balancing of
forces.  We first discuss the former, then add the latter to the analysis.

\subsection{Competing potential energies}\label{sec:com-energy}
We begin by assuming that all five tiles are mechanically stable at any
pressure, meaning that any tile that is generated by random agitations under
centrifugal pressure $\hat{p}(\hat{r})$ remains in place when the agitations
stop abruptly.  In Eqs.~(\ref{eq:5}) the exponents on the left are
\begin{eqnarray}\label{eq:40} 
K_m=\hat{p}(\hat{r})\pm\hat{g},
\end{eqnarray}
where the plus (minus) sign applies to $m=1,4~(2,3)$.
Their physical solution becomes
\begin{eqnarray}\label{eq:41} 
& w_1=w_4=e^{\hat{g}}\left[\sqrt{\sinh^2\hat{g}+e^{2\hat{p}}}+\sinh\hat{g}\right], \nonumber \\
& w_2=w_3=e^{-\hat{g}}\left[\sqrt{\sinh^2\hat{g}+e^{2\hat{p}}}-\sinh\hat{g}\right],
\end{eqnarray}
The tile population densities inferred from (\ref{eq:4}}) are
\begin{eqnarray}\label{eq:42} 
& \bar{N}_1=w_1\bar{N}_4=\frac{w_1w_2}{(1+w_1)(w_1+w_2+2w_1w_2)}, \nonumber \\
& \bar{N}_2=w_2\bar{N}_3=\frac{w_1w_2}{(1+w_2)(w_1+w_2+2w_1w_2)}, \nonumber \\
& \bar{N}_v\doteq 1-\sum_{m=1}^4\bar{N}_m=\frac{2w_1w_2}{w_1+w_2+2w_1w_2}.
\end{eqnarray}

In Fig.~\ref{fig:scen4nmv} we show these population densities plotted versus
$\hat{p}/\hat{g}$ for disks of two different (intermediate) weights.  What
strikes the eye is the extent to which the results of configurational
statistics already reflect the precarious mechanical stability of four of the
five tiles.

\begin{figure}[b]
  \begin{center}
 \includegraphics[width=53mm]{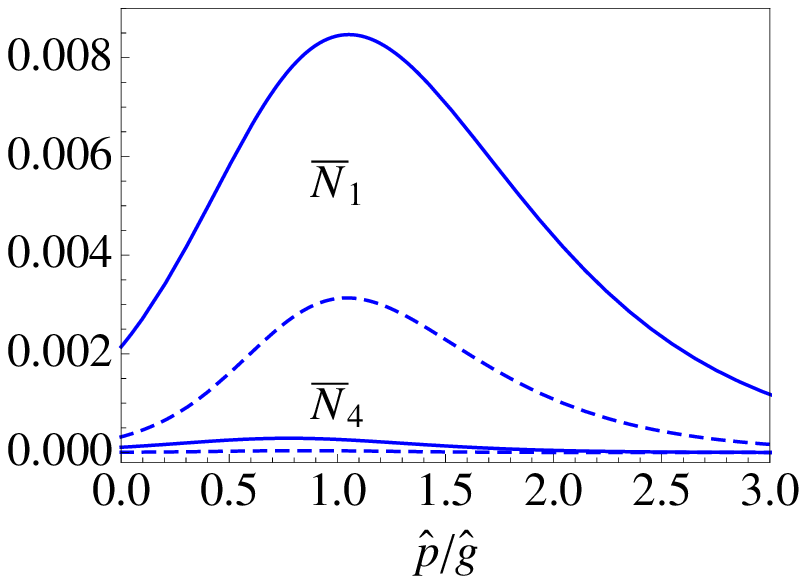}\hspace{2mm}%
 \includegraphics[width=51mm]{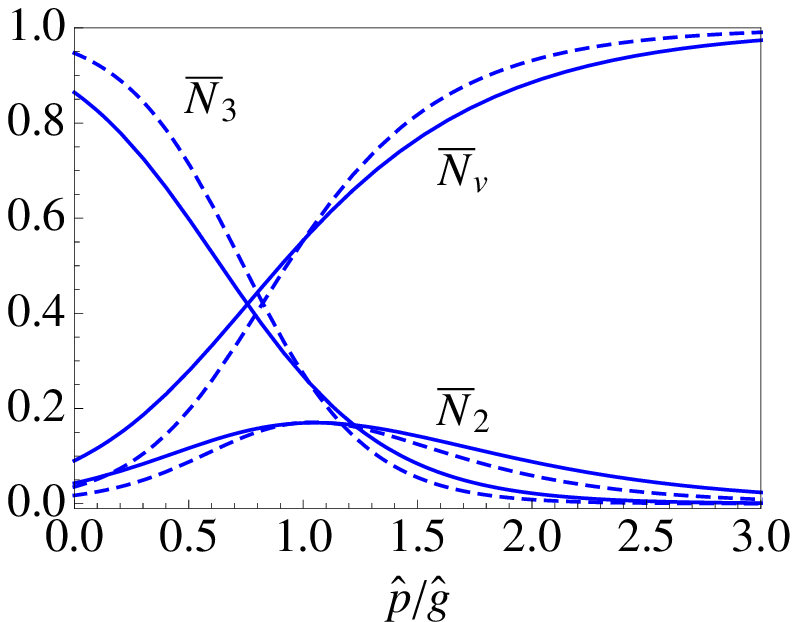}
\end{center}
\caption{Tile population densities (\ref{eq:42}) for $\hat{g}=1.5$ (solid) and
  $\hat{g}=2$ (dashed). Note the different vertical scales left and right.}
  \label{fig:scen4nmv}
\end{figure}

Tiles \textsf{4}, which are mechanically unstable at any pressure, are hardly
present (less than one in a thousand).  Tiles \textsf{1}, which are unstable
at $\hat{p}/\hat{g}<2/3$, are represented by less than half a percent in that
pressure range.  At higher pressure, where their stability becomes
conditional, their representation remains below one percent.

Tiles \textsf{v} dominate at high pressure but their population density dips
down by a factor of ten at low pressure, where they become unstable.  Tiles
\textsf{3} with complementary stability attributes, dominate at low pressure
and virtually disappear at high pressure.

When we increase the weight of the disks by a moderate amount (from
$\hat{g}=1.5$ to $\hat{g}=2$) the population densities of tiles \textsf{1} and
\textsf{4} experience another significant drop whereas those of tiles
\textsf{v} and \textsf{3} reflect their complementary abundance and scarcity
in sharper relief.  The conversion between tiles \textsf{v} and \textsf{3} is
predicted by configurational statistics to take place at $\hat{p}/\hat{g}=1$
when the scaled compression work $\hat{p}$ is balanced by the scaled
gravitational work $\hat{g}$.

\subsection{Competing forces}\label{sec:com-force}
The mechanical instability involving tiles \textsf{v} and \textsf{3} produces
a singularity at $\hat{p}/\hat{g}=\frac{1}{3}$.  Below that threshold value
all tiles \textsf{v} spontaneously collapse into tiles \textsf{3} due to
insufficient centrifugal pressure.  This instability is not reflected in the
results of Fig.~\ref{fig:scen4nmv}.

The competing potential energies come into play while the system is undergoing
random agitations and the competing forces that cause mechanical instabilities
when the agitations stop and the system is jammed.  Configurational statistics
naturally accounts for the former.  The latter must be imposed in an ad-hoc
fashion.

In the current context we proceed by dropping tiles tiles \textsf{1} and
\textsf{4} from consideration, knowing that they are ignorable for
$\hat{g}=1.5$ or higher.  In \ref{sec:scen-iv} we have determined
that at $\hat{p}/\hat{g}\leq\frac{1}{3}$ only tiles \textsf{3} exist and have
estimated that at $\hat{p}/\hat{g}>\frac{1}{3}$ only a fraction,
\begin{equation}\label{eq:34} 
R_g(\hat{p}/\hat{g})=\frac{\sigma/2d}{\sqrt{1+\left(\frac{4d}{\sigma}\frac{\hat{p}}{\hat{g}}\right)^2}}.
\end{equation}
of tiles \textsf{3} that are generated survive jamming by centrifugal
pressure.

For the population density of tiles \textsf{3}, corrected to account for their
conditional mechanical stability, we thus write
\begin{equation}\label{eq:43} 
\bar{N}_3^{(c)}(\hat{p},\hat{g}) = 
\left\{ \begin{array}{ll} 1, & \hat{p}\leq\frac{1}{3}\hat{g} \\
R_g(\hat{p}/\hat{g}){\displaystyle
  \frac{\bar{N}_3(\hat{p},\hat{g})}{\bar{N}_3(\hat{g}/3,\hat{g})}}, 
& \hat{p}>\frac{1}{3}\hat{g}
\end{array} \right.,
\end{equation}
where $\bar{N}_3(\hat{p},\hat{g})$ comes from (\ref{eq:42}).

This expression takes into account that configurational statistics
underestimates the number of tiles \textsf{3} near the mechanical instability.
The correction has the consequence that the number of tiles \textsf{3} are
overestimated by the same factor at high $\hat{p}$, where configurational
statistics is more accurate.  The side effect is negligible because the number
of tiles \textsf{3} decreases rapidly as $\hat{p}$ increases from the point of
instability.

Contrary to what configurational statistics predicts (see
Fig.~\ref{fig:scen4nmv}), there are no tiles \textsf{v} and \textsf{2} at
$\hat{p}/\hat{g}\leq\frac{1}{3}$.  As the centrifugal pressure decreases
toward the point of mechanical instability, $\hat{p}/\hat{g}=\frac{1}{3}$,
tiles \textsf{v} collapse into tiles \textsf{3} and convert \textsf{2} into
tiles \textsf{3} in the process.  That conversion is best captured in the
following corrected population density of tiles \textsf{2}:
\begin{equation}\label{eq:44} 
{N}_2^{(c)}(\hat{p},\hat{g})=[1-{N}_3^{(c)}(\hat{p},\hat{g})]{N}_2(\hat{p},\hat{g}).
\end{equation}
The corrected population densities of tiles \textsf{v} is then the complement,
\begin{equation}\label{eq:45} 
{N}_v^{(c)}(\hat{p},\hat{g})=1-{N}_2^{(c)}(\hat{p},\hat{g})-{N}_3^{(c)}(\hat{p},\hat{g}).
\end{equation}

By avoiding any direct manipulation of ${N}_v(\hat{p},\hat{g})$ we minimize
the effects at high pressure of our approximate accounting for the mechanical
instability.  In Fig.~\ref{fig:scen4nmv-mi} we show the three population
densities corrected for the presence of the mechanical instability in
comparison with their uncorrected counterparts for two values of $\hat{g}$.

\begin{figure}[!ht]
  \begin{center}
 \includegraphics[width=52mm]{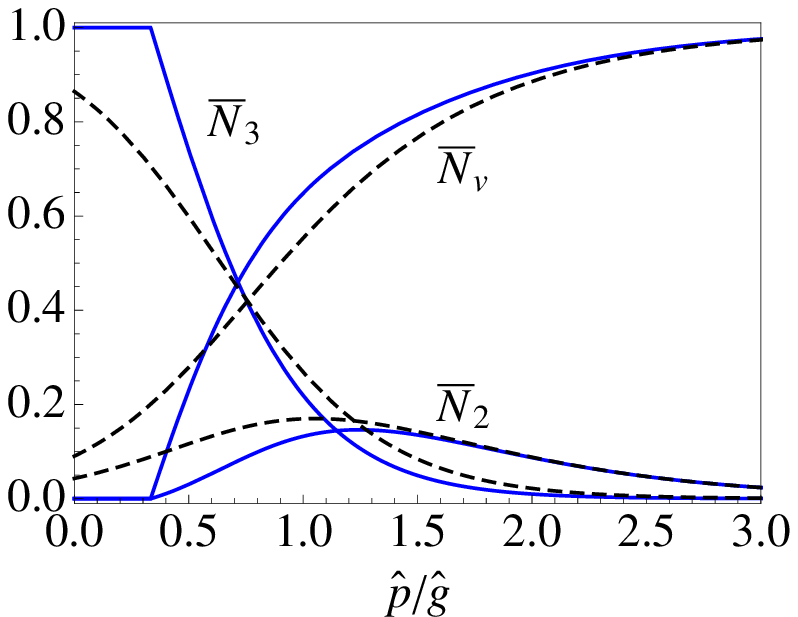}\hspace{2mm}%
 \includegraphics[width=52mm]{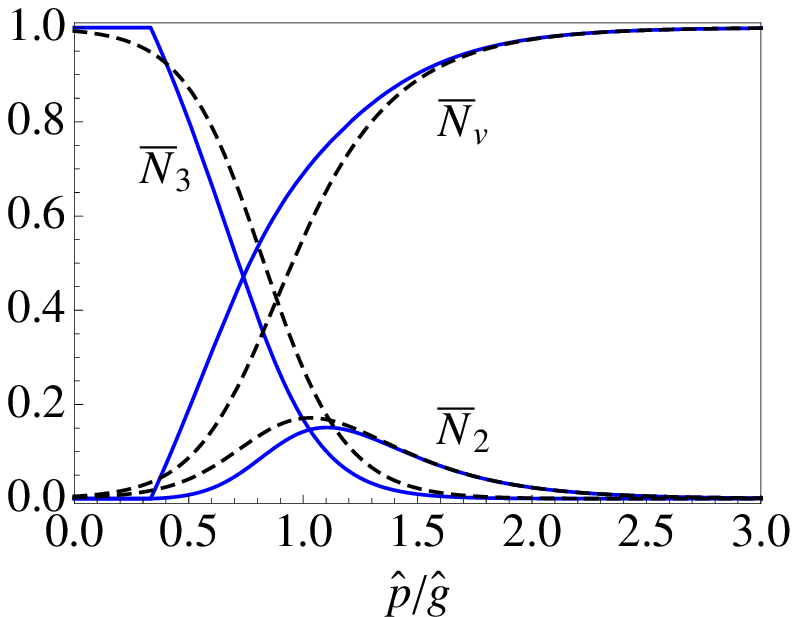}
\end{center}
\caption{Population densities of tiles \textsf{v}, \textsf{2}, and \textsf{3} versus $\hat{p}/\hat{g}$ for $\hat{g}=1.5$ (left) and $\hat{g}=3$ (right). The dashed lines represent the results (\ref{eq:42}). The solid lines account for the mechanical instability at $\hat{p}/\hat{g}=\frac{1}{3}$ via the results (\ref{eq:43})-(\ref{eq:45}).}
  \label{fig:scen4nmv-mi}
\end{figure}

We see that the effects of the mechanical instability do indeed have no
significant impact at high pressure as should be the case.  We also discern
parallel trends (most clearly for $\hat{g}=3$) produced by the competing
potential energies at $\hat{p}/\hat{g}=1$ in the uncorrected results and the
competing forces at $\hat{p}/\hat{g}=\frac{1}{3}$ in the corrected results.

\begin{figure}[t]
  \begin{center}
 \includegraphics[width=52mm]{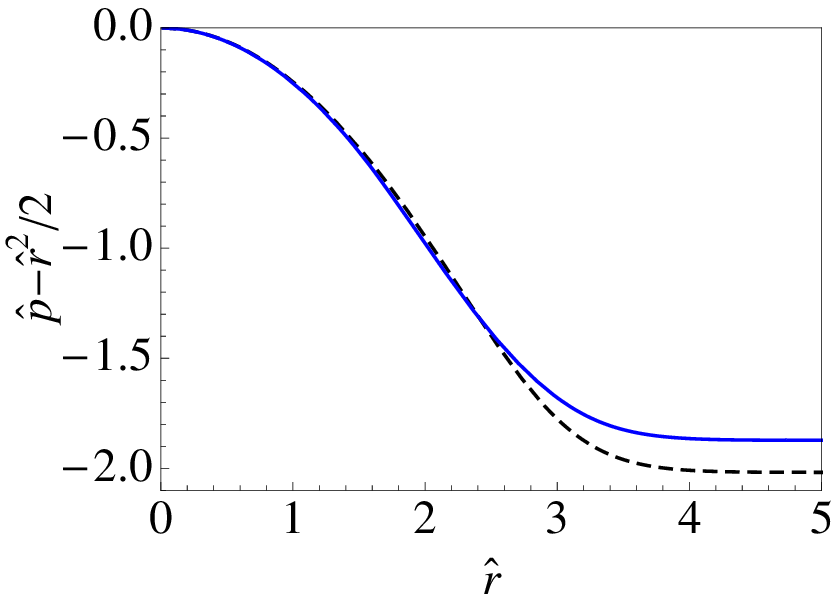}\hspace{2mm}%
 \includegraphics[width=52mm]{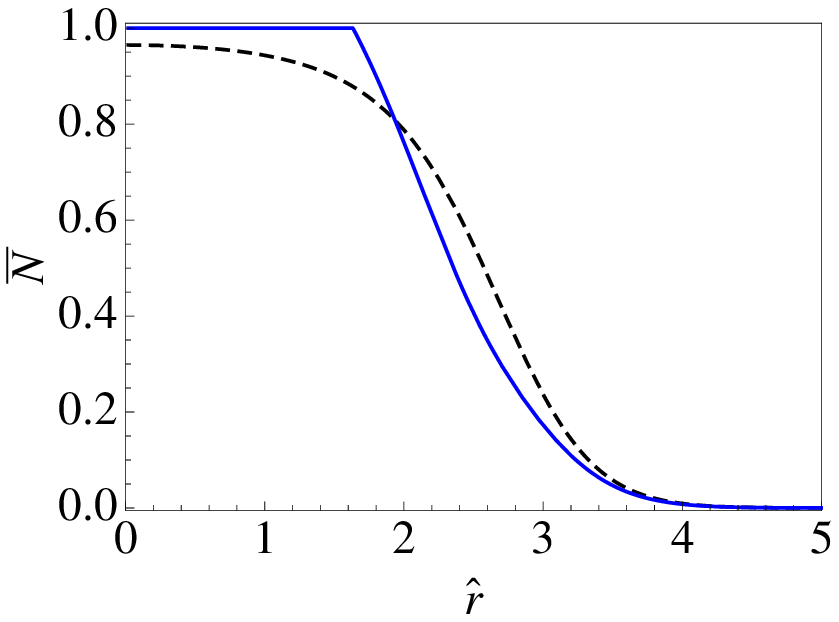}
  \includegraphics[width=52mm]{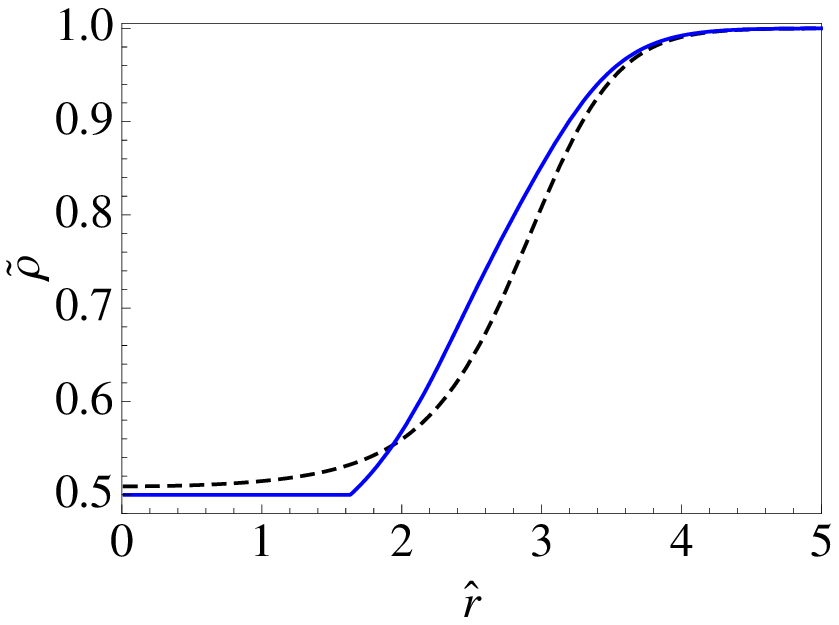}\hspace{2mm}%
  \includegraphics[width=52mm]{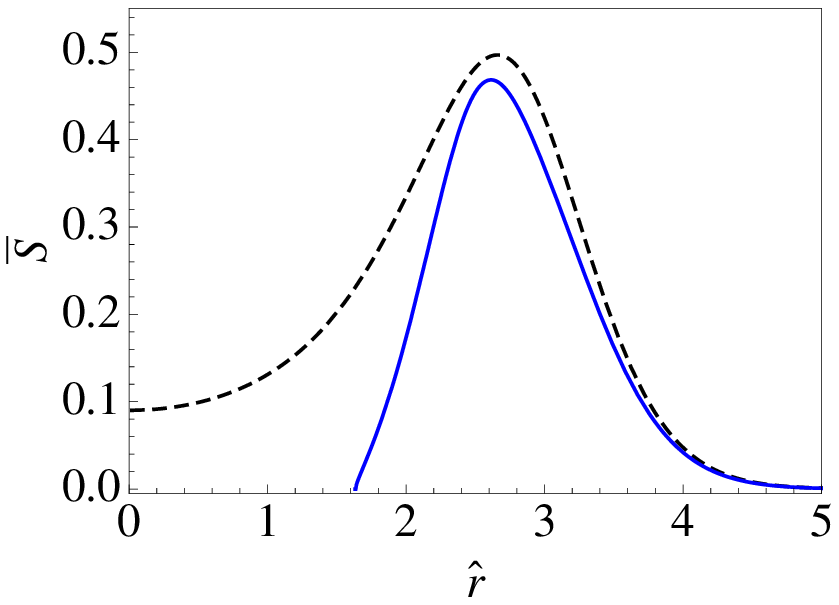}
\end{center}
\caption{Profiles of scaled pressure, scaled population density of particle
  tiles, scaled mass density, and scaled entropy, versus scaled radius in
  scenario (iv) with $\hat{g}=2$. The dashed lines represent the results that
  account only for statistical instabilities. The solid lines also account for
  the mechanical instability at $\hat{p}/\hat{g}=\frac{1}{3}$.}
  \label{fig:scen4uc}
\end{figure}

For the calculation of the four profiles we must use relation (\ref{eq:14})
now relating $\hat{r}(\hat{p},\hat{g})$ to $\bar{N}(\hat{p},\hat{g})$, where
$\bar{N}(\hat{p},\hat{g})$ is the sum of all four population densities in the
uncorrected results and the sum of ${N}_2^{(c)}(\hat{p},\hat{g})$ and
${N}_3^{(c)}(\hat{p},\hat{g})$ in the results corrected for the mechanical
instability.

In Fig.~\ref{fig:scen4uc} we show these four profiles for the case
$\hat{g}=2$.  The dashed curves highlight the effect of competing potential
energies at $\hat{p}/\hat{g}=1$, whereas the solid curves also include the
effects of the competing forces (mechanical instability) at
$\hat{p}/\hat{g}=\frac{1}{3}$.

%
\section{Conclusion}\label{sec:conc}
%
External fields produce characteristic profiles for local pressure, mass
density, entropy density, and other properties of granular matter.  These
profiles may vary sufficiently slowly to be ignorable under many
circumstances, but they are real.  Our study suggests that these profiles are
generated in two stages.

(i) While the random agitations are in operation, configurations of granular
particles are being weighted, in a first selection process, according to
competing potential energies due to contact forces and external forces.

(ii) When the random agitations stop and a jammed microstate is frozen out, a
second selection process eliminates configurations that are mechanically
unstable.

The evidence for these profiles resulting from a two-stage jamming process
pertains, admittedly, to an idealized system of disks confined to a narrow
channel subjected to gravity or centrifuge or both.  However, the scouting for
clues to a deeper understanding of multifaceted processes in complex systems
via studies of idealized scenarios that can be analyzed with some rigor is an
accepted practice with documented success in granular matter studies
\cite{BS06, AB09, BA11, GKLM13, YAB12, AYB13} and elsewhere.

Finally, we wish to comment on a series of theoretical studies by Luck and
Mehta [26-30] that investigate columns of grains in jammed configurations with
trapped void spaces. The methodology is quite different from ours, based on a
stochastic process for transitions between discrete states of individual
grains with different degrees of ordering. The model transition rates depend
on three parameters: (i) the intensity of vibrations akin to our $T_k$,
assumed uniform as in our study; (ii) a scaled activation energy, akin to our
$\hat{p}$, that grows with depth and slows down transitions with increasing
depth; (iii) a local ordering field that controls the presence of trapped
voids at given depth. In our approach, this last task is assumed by the
statistically interacting tiles with activation energies that depend on
depth. It will be interesting to compare the results of the two approaches
once our methodology has been extended to include the kinetics of transitions
between different tile configurations.

\appendix

%
\section{Mechanical instabilities}\label{sec:mech-inst}
%
Here we examine the statistical effects of mechanical instabilities of tiles
under conditions realized in two scenarios.

\subsection{Scenario (iii)}\label{sec:scen-iii}
Two disks touching the same wall, held in place by disks touching the opposite
wall, represent a tile $\mathsf{\bar{1}}$.  It has unconditional mechanical
stability.  A configuration with three or more disks touching the same wall
involves at least one tile $\mathsf{\bar{3}}$.  Its mechanical stability
depends on the strength of the centrifugal force at given local gravitational
pressure.

Configurational statistics predicts expressions (\ref{eq:32}) for the
population densities of tiles $\mathsf{\bar{1}}$ and $\mathsf{\bar{3}}$ under
the assumption that the mechanical stability of all tiles is unconditional.
These tiles are generated by random agitations of given intensity.

For an estimate of the fraction $R_\omega$ of tiles $\mathsf{\bar{3}}$ that
survive jamming, we consider the configuration depicted on the left of
Fig.~\ref{fig:app-scen3}.  Tiles $\mathsf{\bar{3}}$ have a chance to be
generated if three consecutive disks are positioned on the same side of the
channel with the middle one closer to the center (dot-dashed line).

The stabilizing centrifugal force depends on the distance of the middle disk
from the central axis.  The destabilizing force depends on the local pressure
and the relative horizontal position of the middle disk to its two neighbors.
The additional assumption that both outer disks touch the wall has only a mild
impact on the estimate but simplifies the calculation significantly.

\begin{figure}[htb]
  \begin{center}
    \begin{minipage}[!h]{0.4\linewidth}
      {\includegraphics[width=45mm]{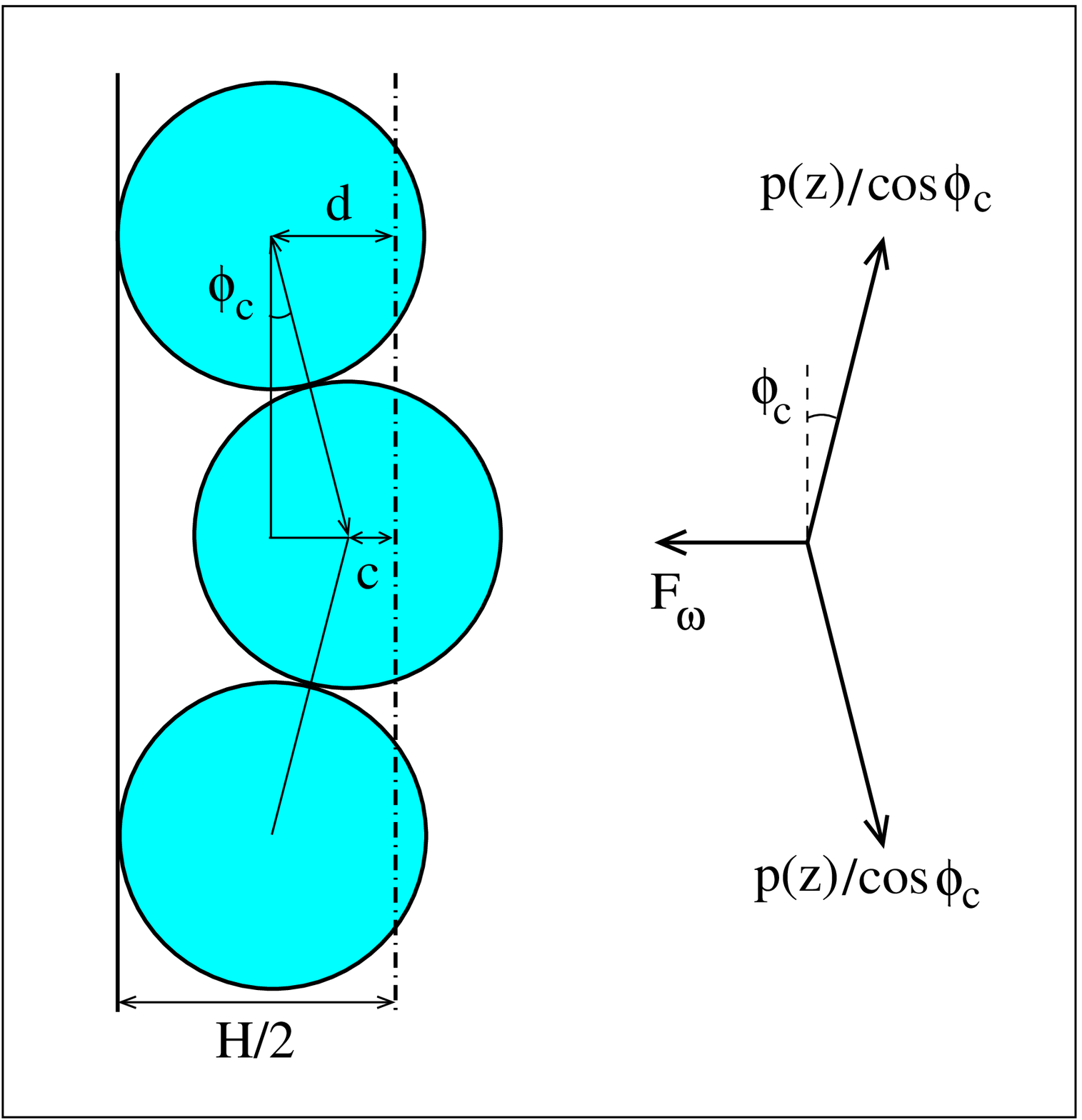}}
    \end{minipage}
    \begin{minipage}[!h]{0.4\linewidth}
      {\vspace{4mm}\includegraphics[width=53mm]{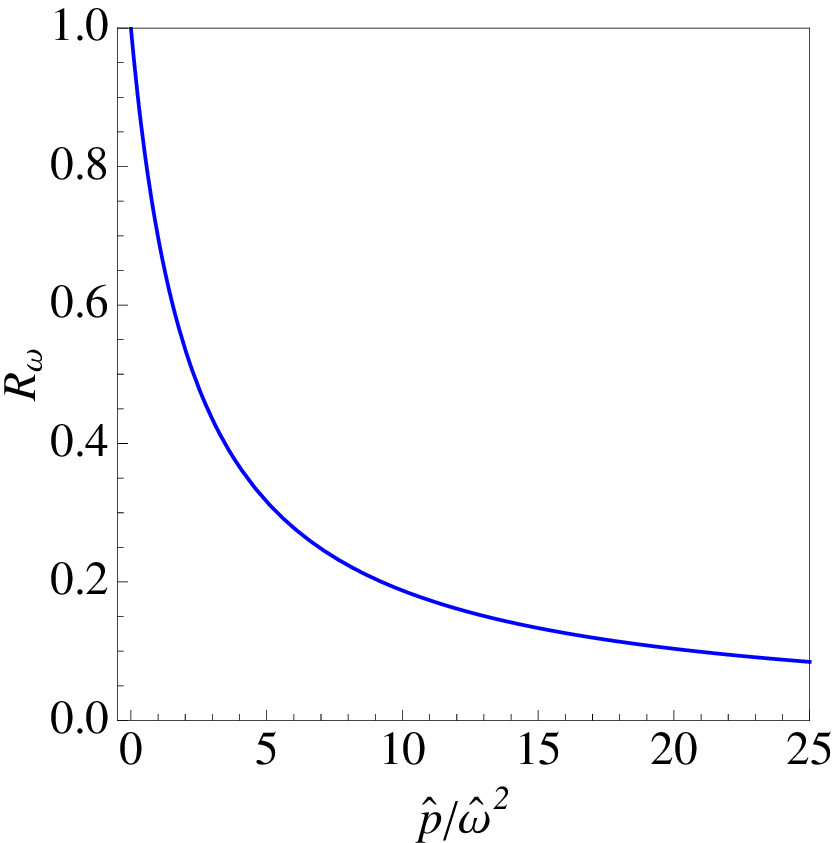}}
    \end{minipage}
  \end{center}

\caption{Three-disk configuration with the stabilizing centrifugal force
  $F_\omega\doteq \mu\omega^2c$ balancing the destabilizing contact forces
  $p(z)/\cos\phi_c$ (left).  Fraction $R_\omega$ of tiles $\mathsf{\bar{3}}$
  predicted to be mechanically stable as a function of
  $\hat{p}/\hat{\omega}^2$ (right).}
  \label{fig:app-scen3}
\end{figure}

The stabilizing and destabilizing forces are in balance if the middle disk is
at position
\begin{equation}\label{eq:a1} 
c=\frac{\hat{p}}{\hat{\omega}^2}d\tan\phi_c\simeq \frac{\hat{p}}{\hat{\omega}^2}d\sin\phi_c,
\end{equation}
where we take into account that the angle $\phi_c$ is small.

The fraction of three-disk configurations shown that lead to tiles
$\mathsf{\bar{3}}$ under jamming is then estimated to be
\begin{equation}\label{eq:a2} 
R_\omega=\frac{d-c}{d},
\end{equation}
which, in conjunction with (\ref{eq:a1}), exhibits the following dependence on
the stabilizing agent $\hat{\omega}$ and the destabilizing agent $\hat{p}$:
\begin{equation}\label{eq:a3} 
R_\omega(\hat{p}/\hat{\omega}^2)=\frac{\sigma/d}{\sigma/d+\hat{p}/\hat{\omega}^2}.
\end{equation}
This function, plotted in Fig.~\ref{fig:app-scen3}, is used in the analysis of
Sec.~\ref{sec:middle}.

\subsection{Scenario (iv)}\label{sec:scen-iv}
Two disks touching the bottom wall, held in place by disks touching the top
wall, represent a tile \textsf{2}.  It has unconditional mechanical stability.
A configuration with three or more disks touching the bottom wall contains at
least one tile \textsf{3}.  Its mechanical stability depends on the strength
of the local centrifugal pressure in relation to the weight of the disk.

Configurational statistics predicts population densities (\ref{eq:42}) under
the assumption that the mechanical stability of all tiles is unconditional.
Here we estimate the fraction $R_g$ of tiles \textsf{3} that survive jamming
by centrifugal pressure.

\begin{figure}[b]
  \begin{center}
    \begin{minipage}[!h]{0.4\linewidth}
      \includegraphics[width=48mm]{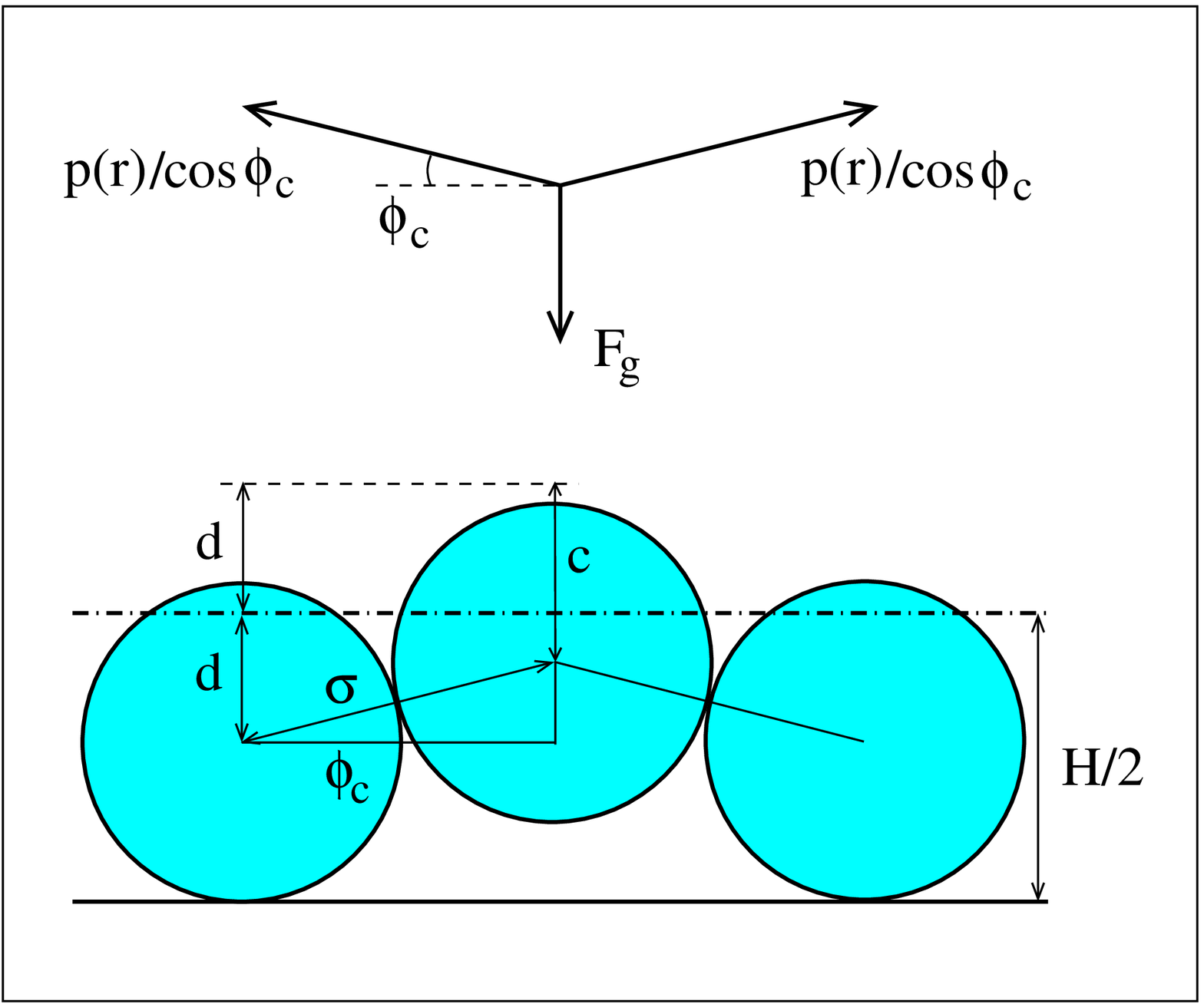}\hspace{2mm}
    \end{minipage}
    \begin{minipage}[!h]{0.4\linewidth}
      \includegraphics[width=53mm]{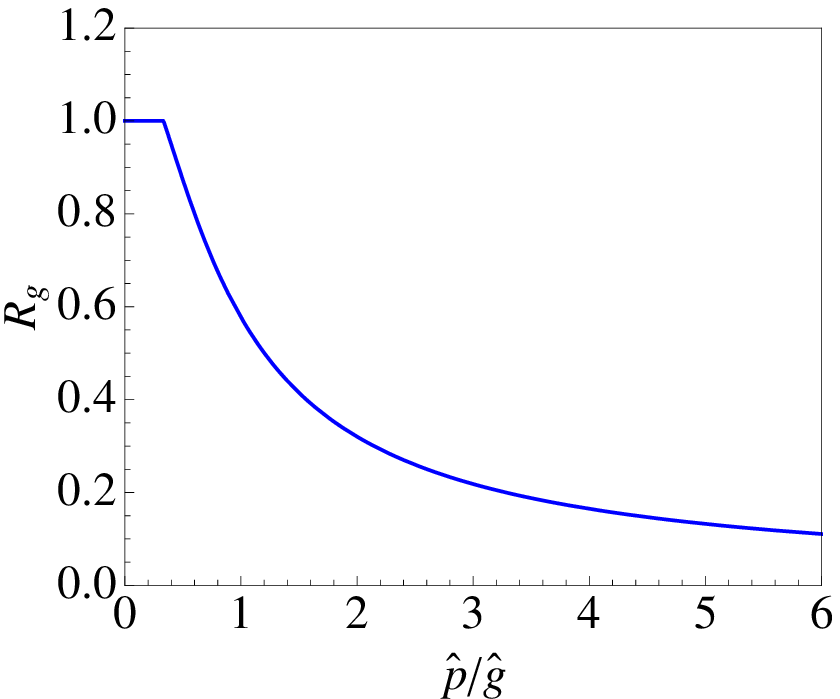}
    \end{minipage}
\end{center}
\caption{Three-disk configuration with the stabilizing gravitational force
  $F_g\doteq \mu g$ in balancing the destabilizing contact forces
  $p(r)/\cos\phi_c$ (left).  Fraction $R_g$ of tiles $\mathsf{3}$ predicted to
  be mechanically stable as a function of $\hat{p}/\hat{g}$ (right).}
  \label{fig:app-scen4}
\end{figure}

We consider the configuration depicted on the left of
Fig.~\ref{fig:app-scen4}.  Tiles \textsf{3} have a chance to be generated if
three consecutive disks are positioned near the bottom of the channel with the
middle one furthest from the wall before jamming begins.  The effectiveness of
the stabilizing weight of the middle disk depends on the local pressure and
the relative vertical position of the middle disk relative to its two
neighbors.  As before we argue that the impact of additional assumptions is
ignorable.

The stabilizing and destabilizing forces are in balance if the middle disk is
at position $c$, implying
\begin{equation}\label{eq:a6}
\cot\phi_c=\frac{4d}{\sigma}\frac{\hat{p}}{\hat{g}}.
\end{equation}
The parameter $\hat{p}/\hat{g}$ varies along the channel in a controlled way.
The fraction of three-disk configurations shown that lead to tiles
$\mathsf{3}$ under jamming is
\begin{equation}\label{eq:a8} 
R_g=\frac{2d-c}{2d},
\end{equation}
from which we infer the function
\begin{equation}\label{eq:a9} 
R_g(\hat{p}/\hat{g})=\frac{\sigma/2d}{\sqrt{1+(4d\hat{p}/\sigma\hat{g})^2}}.
\end{equation}
The ratio levels off in a cusp at $\hat{p}/\hat{g}=\frac{1}{3}$ when it
reaches unity as shown in Fig.~\ref{fig:app-scen4}.

%
%

\section*{References}



\begin{thebibliography}{100}

\bibitem{Fren11}
D. Frenkel in \emph{Understanding soft condensed matter via modeling and computation}, W. Hu and A.-C. Shi (Eds.), World Scientific, Singapore (2011).

\bibitem{OLL+02}
C. O'Hern, S. A. Langer, A. J. Liu, and S. R. Nagel, 
Phys. Rev. Lett. \textbf{88}, 075507 (2002).

\bibitem{MSLB07}
T. S. Majumdar, M. Sperl, S. Luding, and R. P. Behringer,
Phys. Rev. Lett. \textbf{98}, 058001 (2007).

\bibitem{Head09}
D. A. Head, Phys. Rev. Lett. \textbf{102}, 138001 (2009).

\bibitem{SNRS07}
M. Schr\"oter, S. N\"agle, C. Radin, and H. L. Swinney, EPL, \textbf{78}, 44004 (2007).

\bibitem{CCPZ12}
P. Charbonneau, E. I. Corwin, G. Parisi, and F. Zamponi,
Phys. Rev. Lett. \textbf{109}, 205501 (2012).

\bibitem{SWM08}
C. Song, P. Wang, and H. A. Makse,
Nature \textbf{453}, 625 (2008).

\bibitem{BSWM08}
C. Briscoe, C. Song, P. Wang, and H. A. Makse,
Phys. Rev. Lett. \textbf{101}, 188001 (2008).

\bibitem{BSWM10}
C. Briscoe, C. Song, P, Wang, and H. A. Makse,
Physica A \textbf{389}, 3978 (2010).

\bibitem{JM10}
Y. Jin and H. A. Makse,
Physica A \textbf{389}, 5362 (2010).

\bibitem{WSJM10}
P. Wang, C. Song, Y. Jin, and H. A. Makse,
arXiv:0808.2196.

\bibitem{CPNC11}
M. P. Ciamarra, R. Pastore, M. Nicodemi, and A. Coniglio,
Phys. Rev. E \textbf{84}, 041308 (2011).

\bibitem{MSJ+10}
S. Meyer, C. Song, Y. Jin, K. Wang, and H. A. Makse,
Physica A \textbf{389}, 5137 (2010).

\bibitem{LS90}
B. D. Lubachevsky and F. H. Stillinger,
J. Stat. Phys. \textbf{60}, 561 (1990).

\bibitem{UKW05}
T. Unger, J. Kertesz, and D. E. Wolf,
Phys. Rev. Lett. \textbf{94}, 178001 (2005).

\bibitem{ZMSB10}
J. Zhang, T. S. Majumdar, M. Sperl, and R. P. Behringer,
Soft  Matter \textbf{6}, 2982 (2010).

\bibitem{DW09}
K. W. Desmond and E. R. Weeks, Phys. Rev. E \textbf{80}, 051305 (2009).

\bibitem{GBOS09}
G.-J. Gao, J. Blawzdziewicz, C. S. O'Hern, and M. Shattuck, 
Phys. Rev. E \textbf{80}, 061304 (2009).

\bibitem{ABO+12}
S. S. Ashwin, J. Blawzdziewicz, C. S. O'Hern, and M. D. Shattuck,
Phys. Rev. E \textbf{85}, 061307 (2012).

\bibitem{BS06}
R. K. Bowles and I. Saika-Voivod,
Phys. Rev. E \textbf{73}, 011503 (2006).

\bibitem{AB09}
S. S. Ashwin and R. K. Bowles,
Phys. Rev. Lett. \textbf{102}, 235701 (2009).

\bibitem{BA11}
R. K. Bowles and S. S. Ashwin,
Phys. Rev. E \textbf{83}, 031302 (2011).

\bibitem{GKLM13}
N. Gundlach, M. Karbach, D. Liu, and G. M\"uller,
J. Stat. Mech. \textbf{P04018} (2013).

\bibitem{YAB12}
M. Z. Yamchi, S. S. Ashwin, and R. K. Bowles,
Phys. Rev. Lett. \textbf{109}, 225701 (2012).

\bibitem{AYB13}
S. S. Ashwin, M. Z. Yamchi, and R. K. Bowles,
Phys. Rev. Lett. \textbf{110} 145701 (2013).

\bibitem{SLM01} P. F. Stadler, J. M. Luck, and
  A. Mehta, Eur. Phys. Lett. \textbf{57}, 46 (2001).

\bibitem{LM03} J. M. Luck and
  A. Mehta, Eur. Phys. J. B \textbf{35}, 399 (2003).

\bibitem{ML03} A. Mehta and
  J. M. Luck, J.Phys. A: Math. Gen. \textbf{36}, L365 (2003).

\bibitem{LM07} J. M. Luck and
  A. Mehta, Eur. Phys. J. B \textbf{57}, 429 (2007).

\bibitem{LM10} J. M. Luck and
  A. Mehta, Euro. Phys. J. B \textbf{77}, 505 (2010).

\bibitem{Hald91a}
F. D. M. Haldane, Phys. Rev. Lett. \textbf{67}, 937 (1991).

\bibitem{Wu94}
Y.-S. Wu, Phys. Rev. Lett. \textbf{73}, 922 (1994).

\bibitem{Isak94}
S. B. Isakov, Phys. Rev. Lett. \textbf{73}, 2150 (1994); Mod. Phys. Lett. B \textbf{8}, 319 (1994).

\bibitem{Anghel}
D.-V. Anghel, J. Phys. A \textbf{40}, F1013 (2007);
Europhys. Lett. \textbf{87}, 60009 (2009).

\bibitem{LVP+08}
P. Lu, J. Vanasse, C. Piecuch, M. Karbach, and G. M{\"u}ller, J. Phys. A \textbf{41}, 265003 (2008).

\bibitem{copic}
D. Liu, P. Lu, G. M\"uller, and M. Karbach,
Phys. Rev. E \textbf{84}, 021136 (2011).

\bibitem{picnnn}
P. Lu, D. Liu, G. M\"uller, and M. Karbach
Condens. Matter Phys. \textbf{15}, 13001 (2012).

\bibitem{pichs}
D. Liu, J. Vanasse, G. M\"uller, and M. Karbach,
Phys. Rev. E \textbf{85}, 011144 (2012).

\bibitem{EO89}
S. F. Edwards and R. B. S. Oakeshott, Physica A \textbf{157}, 1080 (1989).

\bibitem{ME89}
A. Mehta and S. F. Edwards, Physica A \textbf{157}, 1091 (1989).

\end{thebibliography}
\end{document}